\documentclass[10pt， conference， letterpaper]{ IEEEtran}
\IEEEoverridecommandlockouts

\usepackage{cite}
\usepackage{amsmath,amssymb,amsfonts}
\usepackage{algorithm}
\usepackage{algorithmicx}
\usepackage{graphicx}
\usepackage{algpseudocode}
\usepackage{textcomp}
\usepackage{xcolor}
\usepackage{subfig}
\usepackage{makecell}
\usepackage{multirow}
\usepackage{multicol} 
\usepackage{float}
\usepackage{booktabs}
\usepackage{subcaption}
\usepackage{caption}
\usepackage{xspace}
\usepackage{ragged2e}
\usepackage{booktabs}
\usepackage{multirow}
\usepackage{enumitem}

\usepackage[top=1.45cm, bottom=1.45cm, left=1.50cm, right=1.50cm]{geometry}

\def\eg{\textit{e.g.}\xspace}
\def\ie{\textit{i.e.}}

\newcommand{\name}{\emph{Tru-RM}\xspace}
\newcommand{\MA}{\emph{AFD}\xspace}
\newcommand{\MB}{\emph{FPE}\xspace}
\newcommand{\MC}{\emph{PTN}\xspace}

\def\BibTeX{{\rm B\kern-.05em{\sc i\kern-.025em b}\kern-.08em
    T\kern-.1667em\lower.7ex\hbox{E}\kern-.125emX}}
\begin{document}

\title{Adaptive Attribute-Decoupled Encryption for\\ Trusted Respiratory Monitoring in\\ Resource-Limited Consumer Healthcare}

\author{Xinyu Li\textsuperscript{$1$}, Jinyang Huang\textsuperscript{$1\ast$}, Feng-Qi Cui, Meng Wang, \\Peng Zhao, Meng Li, Dan Guo, and Meng Wang

\thanks{Xinyu Li, Jinyang Huang, Feng-Qi Cui, Meng Wang, Peng Zhao, Meng Li, Dan Guo, and Meng Wang, are with the Anhui Province Key Laboratory of Affective Computing and Advanced Intelligence Machine, School of Computer and Information, Hefei University of Technology, Hefei, 230601, China.}
\thanks{Feng-Qi Cui is with the Institute of Advanced Technology, University of Science and Technology of China, Hefei 230026, China.}
}

\maketitle

\begin{abstract}
Respiratory monitoring is an extremely important task in modern medical services. Due to its significant advantages, \eg, non-contact, radar-based respiratory monitoring has attracted widespread attention from both academia and industry. Unfortunately, though it can achieve high monitoring accuracy, consumer electronics-grade radar data inevitably contains User-sensitive Identity Information (USI), which may be maliciously used and further lead to privacy leakage. To track these challenges, by variational mode decomposition (VMD) and adversarial loss-based encryption, we propose a novel Trusted Respiratory Monitoring paradigm, \name, to perform automated respiratory monitoring through radio signals while effectively anonymizing USI. The key enablers of \name are Attribute Feature Decoupling (\MA), Flexible Perturbation Encryptor (\MB), and robust Perturbation Tolerable Network (\MC) used for attribute decomposition, identity encryption, and perturbed respiratory monitoring, respectively. Specifically, \MA is designed to decompose the raw radar signals into the universal respiratory component, the personal difference component, and other unrelated components. Then, by using large noise to drown out the other unrelated components, and the phase noise algorithm with a learning intensity parameter to eliminate USI in the personal difference component, \MB is designed to achieve complete user identity information encryption without affecting respiratory features. Finally, by designing the transferred generalized domain-independent network, \MC is employed to accurately detect respiration when waveforms change significantly. Extensive experiments based on various detection distances, respiratory patterns, and durations demonstrate the superior performance of \name on strong anonymity of USI, and high detection accuracy of perturbed respiratory waveforms.

\end{abstract}

\section{Introduction}
\IEEEPARstart{W}{ith} the rapid development of smart healthcare technologies in consumer electronics, collaborative health monitoring across multiple devices has become increasingly feasible \cite{10414999}. The respiratory status, as a fundamental physiological parameter for daily health monitoring \cite{10632107, 11175575, 10804189, zhao2024wi} and medical diagnosis \cite{10942423}, plays a crucial role in collaborative health monitoring systems. Due to various advantages, \eg, non-contact, miniaturization caused by the rapid development of consumer electronics technology, and applicability for low-light scenarios, millimeter-wave (mmWave) radar has attracted considerable attention from both academia and industry for respiratory monitoring in recent years \cite{liu2025multi, 10745119}.

\begin{figure}[t]
\centerline{\includegraphics[width=0.48\textwidth]{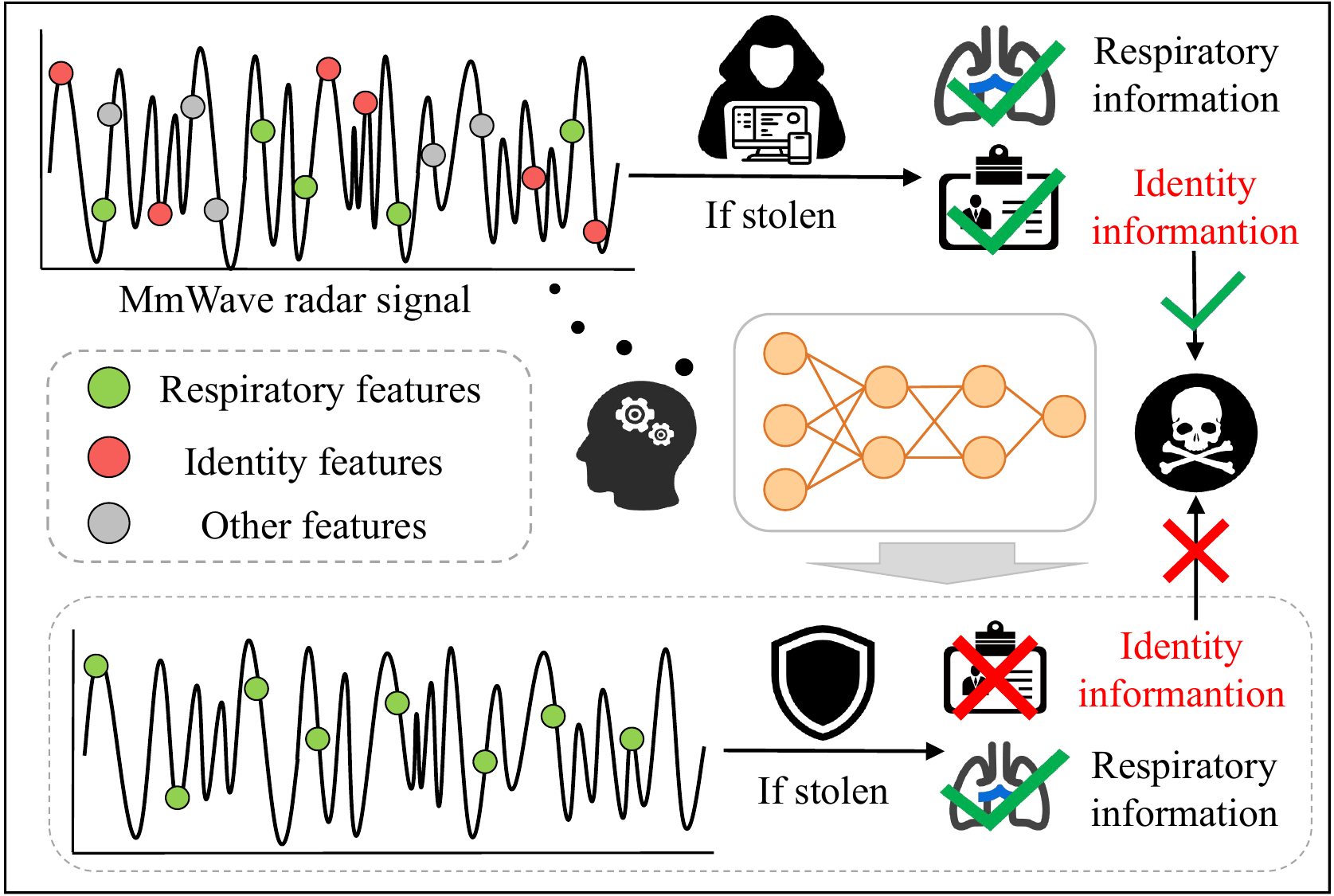}}
\caption{Identity recognition based on mmWave radar.}  
\label{feature}
\vspace{-0.2in}
\end{figure}

Unfortunately, as illustrated in Fig.~\ref{feature}, consumer electronics-grade radar data inevitably contain User-sensitive Identity Information (USI) \cite{9864117, yang2020mu}, \eg, individual-specific respiratory patterns and rhythms captured during respiratory monitoring, which can be exploited for user identity recognition. This inherent characteristic poses significant internal privacy leakage risks in collaborative health monitoring systems, \eg, when multiple sensing devices exchange local information through device-to-device communication \cite{10414999}, the legitimate radar data containing respiratory features sharing process instead opens the channel for the spread of USI. Therefore, it is necessary to introduce an effective and lightweight privacy protection paradigm \cite{liu2025gsfl, 10856209} for radar data to anonymize identity information, thereby ensuring privacy and security in device-to-device communication in collaborative health monitoring systems.

By deploying reflective tags or backscatter arrays in the environment \cite{10931026, shenoy2022rf}, pioneering works attempted to tamper with human presence detection or signal signatures. However, these methods typically depended on specific environmental setups and additional hardware, leading to high costs and complex deployments. More importantly, the respiratory information contained therein is inevitably distorted. Other state-of-the-art works, \eg, \cite{liu2023application, liu2025multi}, introduced inter-frame phase fluctuations and dimensional perturbations into radar signals to enhance privacy level. Nevertheless, such approaches relied on global perturbations to mitigate privacy risks, which also affects the completeness of the physiological information contained in the signals. Although existing works have achieved privacy protection, they often come at the cost of monitoring performance, which contradicts the original goal of collaborative health monitoring to pursue high-precision sensing. Apparently, achieving a balance Utility-Privacy trade-off is extremely difficult due to the inherent conflict between detection performance and privacy protection. The main challenges are demonstrated as follows:

\begin{itemize}
\item \textbf{\emph{Attribute Feature Decoupling:}} In respiratory monitoring, the signals captured by mmWave radar inherently contain both respiratory information and USI, which are highly entangled due to inherent physiological correlations. This intrinsic correlation poses a significant challenge for effectively decoupling them.

\item \textbf{\emph{Flexible Perturbation Encryptor:}} Due to inherent physiological correlations, any perturbation targeting identity features inevitably affects the integrity and availability of respiratory features. Thus, how to flexibly perturb identity features while preserving key respiratory features becomes a question worthy of consideration.

\item \textbf{\emph{Noised Respiratory Monitoring:}} After the perturbation encryption of identity features, the radar signals are contaminated by noise, which inevitably changes characteristics of the signal, \eg, the waveform structures and envelope features, thereby significantly increasing the difficulty of extracting precise and effective respiratory information. Therefore, it is challenging to accurately obtain respiratory features from the noised signals.
\end{itemize}

To address these challenges, in this paper, we propose a trusted respiratory monitoring paradigm, \name, which enables automated respiratory monitoring from radio signals that effectively anonymize USI. Instead of relying on specific hardware devices or indiscriminate global perturbations to tamper with signal features, which safeguards privacy by changing the overall signal structure at the expense of degrading sensing performance, \name first achieves flexible and targeted decoupling and encryption of USI based on the progressive signal processing and adversarial loss-based encryption. Furthermore, to cope with the noised signals under significant waveform variations introduced by encryption perturbations, a robust respiratory feature extraction network based on the spectral distribution alignment is adopted to achieve trusted respiratory monitoring. Compared with state-of-the-art privacy protection methods\cite{liu2025multi}, \name is more secure and practical since it perfectly realizes the balanced trade-off between privacy protection and high-accuracy respiratory monitoring

Specifically, to effectively decouple respiratory features and USI, we design a two-step progressive strategy that combines a respiratory rate-dependent Butterworth bandpass filter and a variational mode decomposition  (VMD) algorithm. Subsequently, targeted noise generated by a key-controlled perturbation encryptor is injected into the USI to achieve effective anonymization. To further process the small amount of identity information contained in the respiratory components, a phase noise algorithm with a learning intensity parameter is designed without impacting respiratory features. Meanwhile, to ensure robust respiratory monitoring even under significant waveform variations introduced by encryption perturbations, we further propose a transferred generalized domain-independent network, which enables accurate extraction of respiratory features from the perturbed signals.

In brief, our contributions can be summarized as follows:
\begin{itemize}
\item To the best of our knowledge, this paper is the first attempt to perform automated respiratory monitoring through radio signals that effectively anonymizes USI for the privacy-friendly trusted respiratory monitoring.
\item By employing the VMD algorithm and the adversarial loss-based encryption algorithm, a novel attribute feature decoupling strategy and the flexible perturbation encryptor are proposed, which effectively overcome the problems posed by indiscriminate global perturbations that often compromise signal integrity and usability.
\item To deal with the noised signals with different waveforms caused by various encryption strengths, by referring spectral distribution alignment strategy, a transferred generalized domain-independent network is proposed.
\item Extensive experimental results demonstrate the superior performance of the proposed scheme on strong anonymity of ID, high accuracy of respiratory detection, and significant morphological variability. 
\end{itemize}

The rest of the paper is organized as follows. Sec.~\ref{Related Work} discusses numerous related studies. Then, we describe the \name system design in Sec.~\ref{system}. Implementation, evaluation, and the impacts of various factors on \name performance are presented in Sec.~\ref{Experimental Evaluation}. Finally, we conclude our work in Sec.~\ref{Conclusion}.

\section{Related Work}\label{Related Work}
\subsection{Respiratory Monitoring Methods}
Long-term and continuous respiratory monitoring is critical for the early diagnosis and effective health management of chronic respiratory diseases \cite{10632107}. Traditional respiratory monitoring methods typically rely on direct contact between sensors and patients, determining respiratory status by measuring physical parameters, \eg, chest and abdominal movements \cite{sethuraman2021mywear}, breath sounds, and airflow \cite{chen2022supervised} generated during respiration. However, due to their complex operational procedures and the considerable discomfort caused by prolonged sensor attachment, these methods are unsuitable for continuous daily health monitoring. The emergence of portable devices, \eg, smart watches \cite{10456913}, has extended respiratory monitoring from clinical settings to daily life, enabling continuous tracking of key respiratory parameters, like breathing rate \cite{10632107}. Nevertheless, since these devices require direct skin contact, their measurements are susceptible to variations in wearing position, tightness, and users’ daily activities, leading to low long-term compliance.

Owing to its non-invasive and contact-free characteristics, non-contact respiratory monitoring technology has attracted considerable attention in recent years. Existing research mainly focuses on camera-based monitoring methods \cite{romano2021non, 10511074, 10158936} and wireless sensing-based monitoring methods, \eg, Wi-Fi-based sensing \cite{hu2024m, song2024finersense} and mmWave radar-based sensing \cite{wang2022your, 10160149, 10633871, xiang2022high, ahmed2023machine, kang2024respiration}. Camera-based methods typically rely on cameras to capture subtle chest and abdominal movements for respiratory signal estimation. Although they can achieve relatively high monitoring accuracy, they inevitably raise privacy concerns because they require direct visual acquisition. In contrast, wireless sensing-based methods perceive respiration by analyzing changes in the reflection or scattering of signals caused by human respiration, eliminating the need for visual data and thereby providing inherent privacy advantages. Specifically, mmWave radar has emerged as a research hotspot in wireless sensing for respiratory monitoring due to its robustness against ambient light variations, high spatial resolution, and strong signal penetration capability.

\begin{figure}[t]
\centerline{\includegraphics[width=0.48\textwidth]{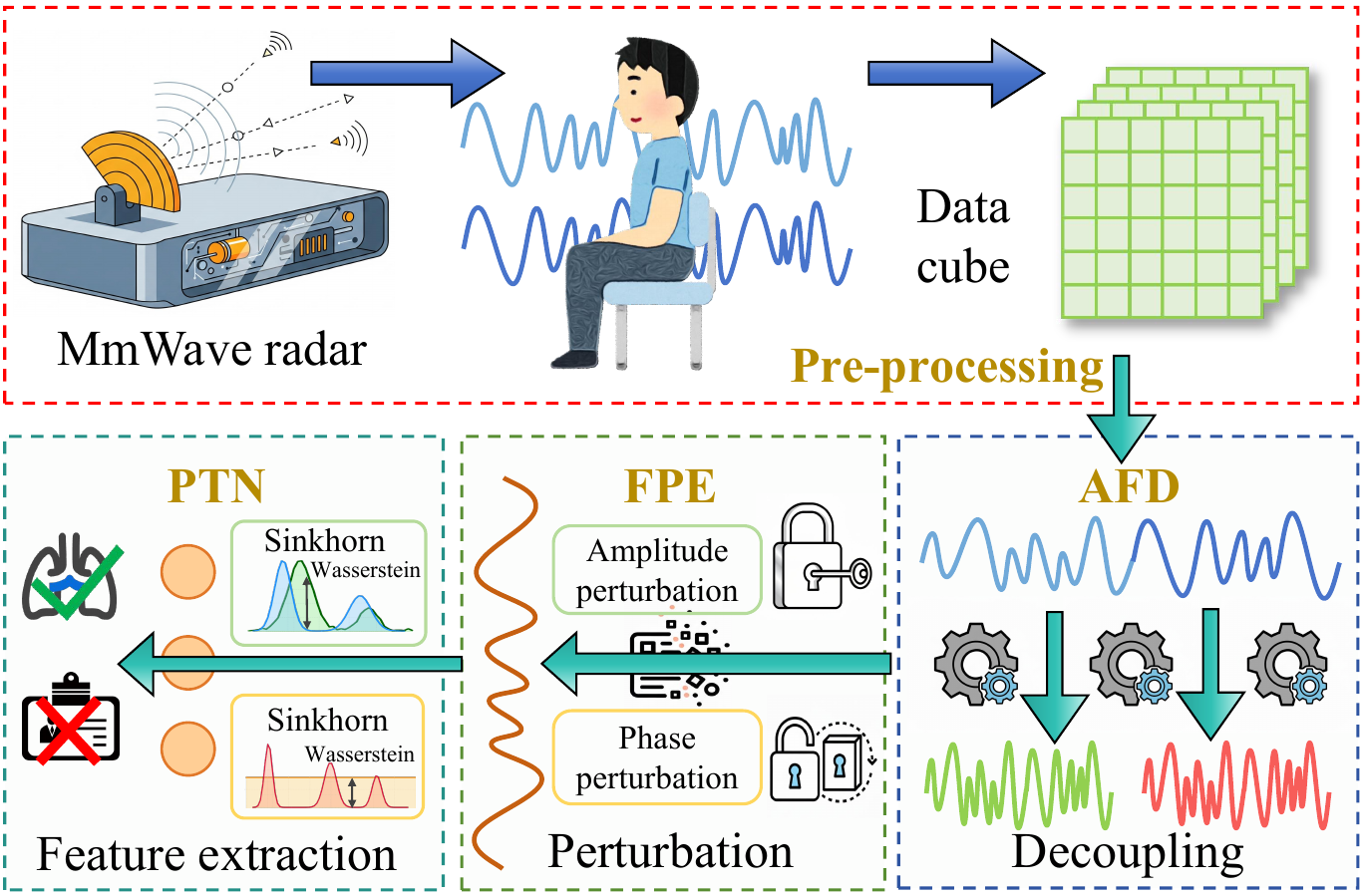}}
\caption{Key processing steps in \name.}
\label{overview}
\vspace{-0.1in}
\end{figure}

\subsection{RF-based Sensing Security and Privacy}
With the widespread application of radio frequency (RF) technology in intelligent sensing, the accompanying security and privacy issues have become increasingly prominent. For instance, sensitive information, including passwords displayed on phones or screens, can be intercepted via RF signals \cite{huang2023phyfinatt, huang2024keystrokesniffer, li2022spiralspy}. Furthermore, mmWave radar and RF identification (RFID) systems were demonstrated to enable non-contact eavesdropping on speech by capturing subtle vibrations of objects \cite{10229095, xu2024mmear}. Users’ location privacy issues cannot be ignored. Through-wall radar technology can reveal the locations and behavior patterns of individuals inside a room \cite{wang2024real, zhu2022twlbr}, thereby increasing potential safety risks. Moreover, physiological signals and gait information captured by mmWave radar were exploited for identity recognition, posing significant privacy risks \cite{zhao2024mn, guo2024millimeter}. Such risks are particularly concerning for sensitive groups, including political figures and individuals with access to confidential information, as they may lead to severe privacy breaches and pose threats to personal safety.

To address privacy leakage concerns, several studies employed physical-layer attacks, utilizing specific devices to interfere with or manipulate the existence or location information of the target \cite{liu2025multi}. For example, RF-Protect \cite{shenoy2022rf} injected a fabricated backscatter signal to mislead potential eavesdroppers by introducing reflectors into the environment, while Radar² \cite{qiu2023radar} enabled users to detect the presence of other mmWave radars and determine their spatial locations using a single mmWave radar. Additionally, pioneer studies focused on customizable signal perturbations to meet personalized privacy requirements. For instance, mmFilter \cite{liu2023application} protected the privacy of individual users through targeted dimensional perturbations, while MuFilter \cite{liu2025multi} extended this approach to multi-user scenarios, enabling broader privacy protection by enabling subspace-level privacy protection policies on demand. However, while these methods provided privacy protection, they often introduced uncontrollable interference to critical sensing features, which inevitably impaired perception tasks and limited their practical applicability in scenarios including medical monitoring and smart home environments.

In contrast to previous studies, we first decouple respiratory features and USI from consumer electronics-grade radar signals, ensuring that perturbations to USI do not affect critical respiratory features. Then, by introducing a key-based amplitude-phase domain perturbation encryptor, we thus achieve targeted encryption of the USI. Additionally, we propose a robust respiratory feature extraction network to enable high-accuracy and trusted respiratory monitoring from the perturbed signals. Compared with global perturbation methods, the proposed scheme ensures both reliable respiratory monitoring and effective privacy preservation by selectively perturbing only towards USI.

\section{System Design}\label{system}

Fig.~\ref{overview} demonstrates the framework of \name. Specifically, \name achieves privacy-friendly trusted respiratory monitoring through four basic modules: Data Pre-processing, Attribute Feature Decoupling (\MA), Flexible Perturbation Encryptor (\MB), and Robust Perturbation Tolerable Network (\MC).

\textbf{\emph{Data Preprocessing:}} The first module is to extract the phase sequence and segment it into fixed-length analysis units. Since the phase is highly sensitive to subtle movements, it preserves sufficient information on both respiration and other micro-movements, thereby facilitating feature extraction.

\textbf{\emph{Attribute Feature Decoupling (AFD):}} This module is designed to decouple the raw radar data to the personal difference component, the universal respiratory component, and the other unrelated components. To achieve this, we design a two-step progressive strategy that combines a respiratory rate-dependent Butterworth bandpass filter and a VMD. By leveraging the distinct intrinsic frequency distributions of these components, this strategy can initially decompose them.

\textbf{\emph{Flexible Perturbation Encryptor (FPE):}} This module aims to achieve flexible and targeted encryption of USI. By using large noise to drown out the other unrelated components, and the phase noise algorithm with a learning intensity parameter to further eliminate USI in the personal difference component, we finally propose a flexible encryption algorithm to conceal USI from both amplitude and phase, effectively retaining respiratory information while encrypting user identity.

\textbf{\emph{Robust Perturbation Tolerable Network (PTN):}} To ensure robust respiratory monitoring even under significant waveform variations introduced by encryption perturbations, by referring to the spectral alignment strategy, we propose a transferred generalized domain-independent network that enables accurate extraction of respiratory features from the perturbed signals.

\subsection{Data Pre-processing} 
The signal received by the mmWave radar consists of reflections from both the target object and the surrounding environment. To separate and extract the target's reflected signals, we first perform Range-FFT on the raw signal along the fast time dimension to convert the signal from the time domain to the frequency domain to obtain the target's distance information. Then, based on the amplitude distribution of the Range-FFT results, the corresponding target range bin is determined, and the signal $x_{tar}$ at this range bin is extracted as the target data.

Since the phase of the signal is highly sensitive to subtle movements, it offers higher precision and reliability in detecting fine-grained human body motions \cite{hao2025detection}. Therefore, by computing the arctangent of its imaginary and real components, we further extract the phase information $\phi_{\text{tar}}$ from $x_{\text{tar}}$:

\begin{equation}\label{arctan}
\phi_{\text{tar}}(t) = \arctan\left( \frac{\operatorname{Im}(x_{\text{tar}}(t))}{\operatorname{Re}(x_{\text{tar}}(t))} \right),
\end{equation}
where $t$ represents different time instances.

To reduce complexity, $\phi_{\text{tar}}$ is divided into $I$ segments. $I$ is calculated by two critical factors, \ie, the segment length and the overlapping region between adjacent segments. The segment length is set such that the duration is larger than one respiratory period, guaranteeing that each segment captures the overall property of respiratory morphology. The overlapping region enables the information to be shared between segments.

\subsection{Attribute Feature Decoupling (\MA)}
In fact, the raw radar signals contain not only respiratory components induced by the periodic chest wall motion during respiration, but also other respiration-independent components arising from other subtle physiological activities, \eg, cardiovascular motion and involuntary micro-movements. Since these components exhibit significant inter-individual differences, both types of components embed USI.

\begin{figure}[t]
\centerline{\includegraphics[width=0.48\textwidth, height=0.23\textheight]{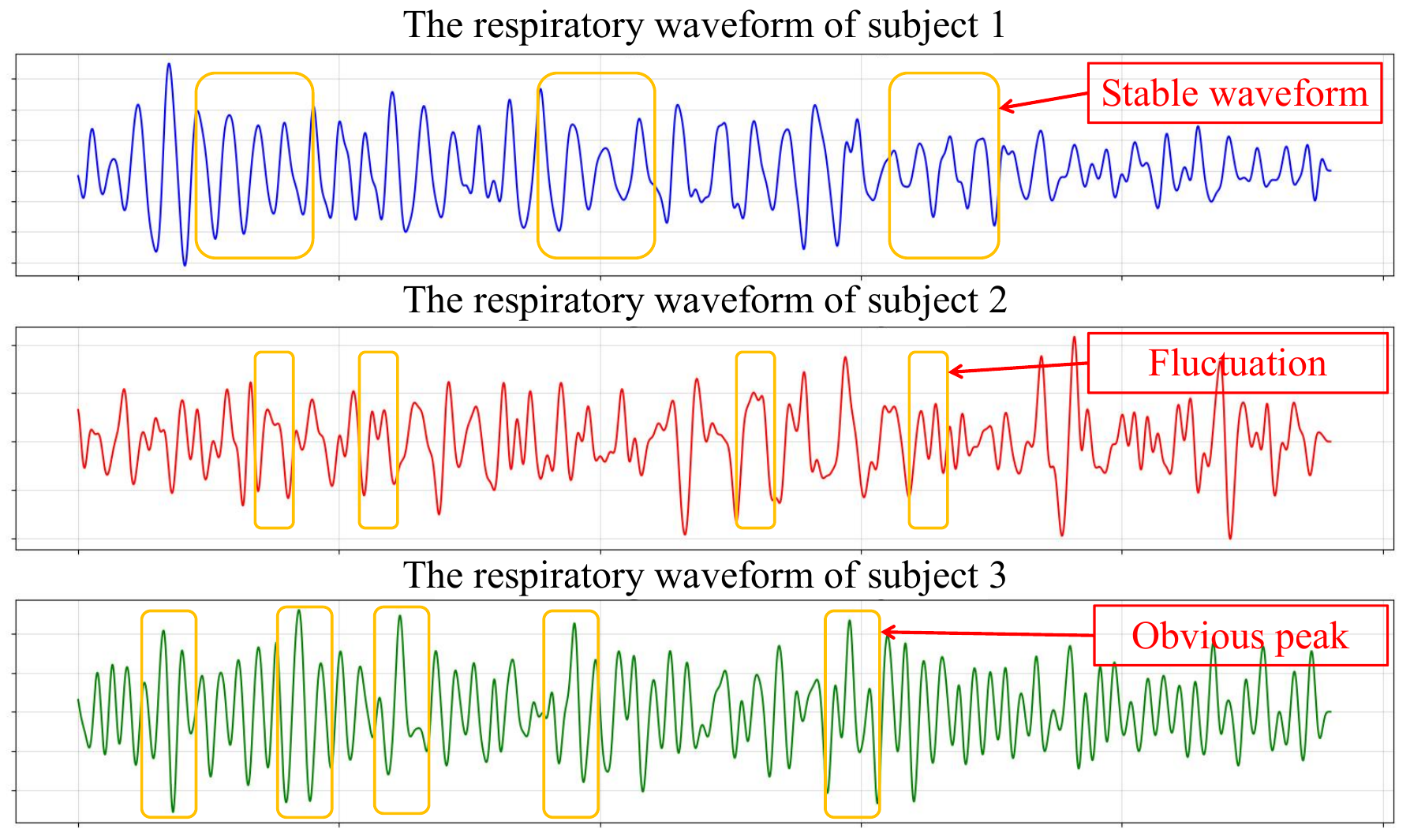}}
\caption{Comparison of respiratory time domain waveforms of different users.} 
\label{respiration_wave}
\vspace{-0.1in}
\end{figure}

To decouple the key respiratory components and the components embedded with USI, we design a two-step progressive strategy, called \MA, that combines a respiratory rate-dependent Butterworth bandpass filter and a VMD algorithm. Firstly, considering that respiratory features are predominantly concentrated within a fixed frequency range of 0.1-0.5 Hz \cite{hao2024mmwave}, a bandpass filter is applied for the raw radar data for preliminarily isolating the respiratory component $x_{re}(t)$ and other unrelated components $x_{ot}(t)$. Nevertheless, due to the unique respiratory patterns of each individual, USI is inevitably embedded in $x_{re}(t)$, primarily manifested in the waveform morphology. As illustrated in Fig.~\ref{respiration_wave}, respiratory waveforms exhibit significant inter-individual differences among different persons in waveform structure. In contrast, the respiratory waveforms of the same person recorded at different times remain highly consistent, demonstrating good stability and reproducibility. Thus, the respiration data can serve as an important tool for user authentication, leading to the identity information leaking when respiration sensor data is captured.

Then, to further decouple the USI embedded in $x_{re}(t)$, based on the adaptive decomposition capability and effective spectral concentration of VMD, we employ this algorithm to decompose $x_{re}(t)$ into the personal difference component $x_{pd}(t)$ and the universal respiratory component $x_{ure}(t)$:
\begin{equation}\label{respiratory}
x_{re}\left ( t \right ) =\underset{x_{ure}}{\underbrace{u_1\left ( t \right )}}    + \sum_{k=2}^{K}\underset{x_{pd}}{\underbrace{u_k\left ( t \right )}  },
\end{equation}
where $K$ represents the number of modes for VMD decomposition. $u_1(t)$ denotes the universal respiratory component $x_{ure}(t)$, which represents the primary component of the respiratory signals. $\sum_{k=2}^{K} u_k(t)$ denotes the personal difference component $x_{pd}(t)$, which contains less respiratory information but more identity-related information due to its relatively weak volatility. In particular, all components except for the universal respiratory component $x_{ure}(t)$ are considered as USI components of signals. This is because not only does the personal difference component $x_{pd}(t)$ relate to identity information, but also the other unrelated components $x_{ot}(t)$ contain identity information. The subsequent algorithm has specific research on this issue.

\subsection{Flexible Perturbation Encryptor (\MB)} \label{encryptor}
To achieve flexible and targeted encryption of USI, we propose the \MB, which consists of a key-controlled amplitude perturbation encryptor and a phase noise algorithm with a learning intensity parameter. By targeted injection of noise into the USI, \MB effectively conceals USI while preserving the essential respiratory features.

\subsubsection{Amplitude Perturbation Encryptor} \label{amplitude domain} 
Although the primary features relevant to respiratory monitoring are concentrated within the respiratory frequency band, other unconscious and involuntary individual features (\eg, those induced by heartbeat or subtle micro-movements) often appear in other frequency bands and may contain distinctive personal information, which can be used for user authentication. Therefore, concealing these USI-related features is necessary for ensuring privacy protection.

According to Parseval’s theorem, the total energy of a signal is conserved between the time and frequency domains. When noise is added to a specific frequency band in the frequency domain, it only affects the corresponding component in the time domain without interfering with other bands. This means that the noise added to the frequency band outside the respiratory frequency band will not affect the respiratory characteristics, but can effectively filter out the user identity information outside the respiratory frequency band. Therefore, we first employ the amplitude perturbation encryptor to inject noise into the $x_{ot}$ in the non-respiratory frequency bands to conceal USI in these frequency bands.

Inspired by the Cosine-transform-based chaotic system for encryption \cite{hua2019cosine}, we design a key-controlled sine-cosine-transform-based perturbation noise $N_f$ to adaptively other unrelated components $x_{ot}(t)$, which can be expressed as:
\begin{equation}\label{noise}
N_f\left ( k \right ) =\alpha _f\cdot \sin \left ( \Gamma \left ( k \right )  \right ) \odot \cos \left ( \Gamma \left ( k \right )  \right ) ,
\end{equation}
where $\alpha_f$ denotes the adaptive noise intensity and can be denoted in Eq.~\ref{alpha}, and $k = \{k_0, k_1, \ldots, k_{L-1}\}$ represents a binary key bit vector of length $L$, in which $k_i \in \{0,1\}$. $\Gamma(k)$ is the seed generated by the standard binary-weighted mapping to ensure the uniqueness and security of the encryption perturbation process, which can be represented as follows:

\begin{equation}\label{T}
\Gamma \left ( k \right ) =\sum_{i=0}^{L-1} k_i\cdot 2^i.
\end{equation} 

\begin{figure}[t]
\centerline{\includegraphics[width=0.48\textwidth, height=0.22\textheight]{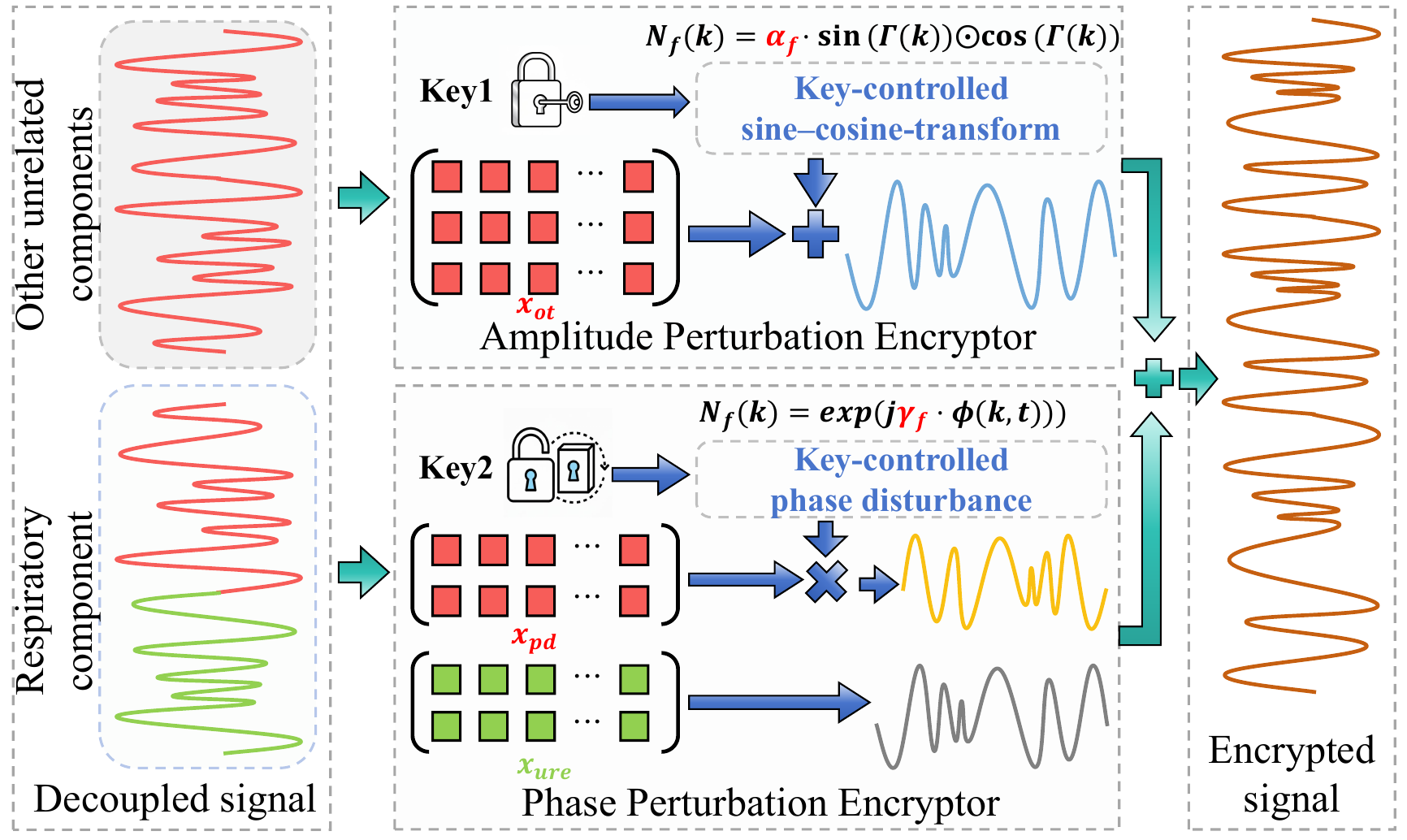}}
\caption{The structure of \MB for perturbation encryption.}  
\label{FPE}
\vspace{-0.1in}
\end{figure}

It is worth noting that, due to variations in signal strength across different frequency bands, fixed-intensity perturbations may result in either excessive or insufficient noise injection. To overcome this limitation, we define $\alpha_f$ as a signal power–based adaptive noise adjustment hyperparameter, where the noise intensity is dynamically scaled in proportion to the local energy of the target frequency band and can be represented as follows:
\begin{equation}\label{alpha}
\alpha_f=\beta \cdot \sqrt{\frac{1}{T} \sum \left | X\left ( f \right )  \right |^2 } ,
\end{equation}
where $\beta$ is a learnable weight parameter optimized via backpropagation of the loss function in Eq.~\ref{loss}. This parameter is proportional to the accuracy of identity recognition, and its primary goal is to maximize the anonymization of USI. $T$ represents the length of the target frequency band, and $X(f)$ denotes the signal at the corresponding frequency $f$.

\begin{figure*}[t]
\centerline{\includegraphics[width=0.95\textwidth]{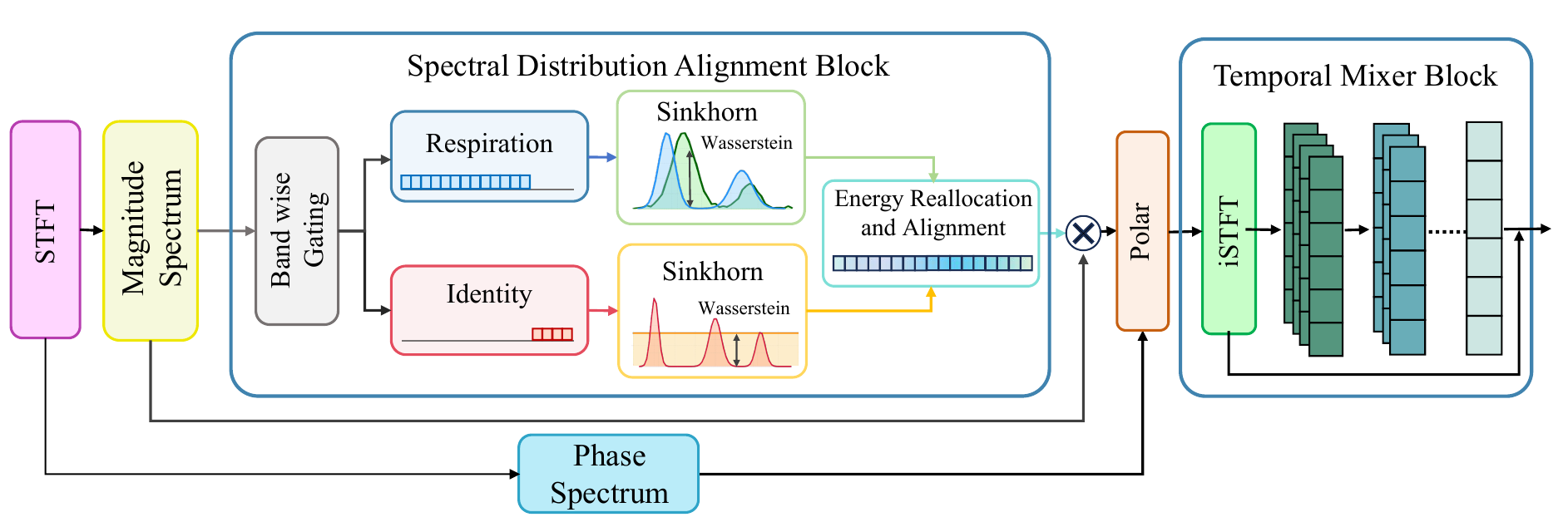}}
\caption{Robust respiratory feature extraction model \MC for perturbed signals}  
\label{network}
\end{figure*}

\subsubsection{Phase Perturbation Encryptor}\label{phase domain}
The amplitude perturbation encryptor is designed to conceal the USI embedded in the other unrelated components $x_{ot}(t)$ that are outside the respiratory frequency band. Nevertheless, the personal difference component $x_{pd}(t)$ in the respiratory frequency band still contains potential USI, manifested as distinctive respiratory waveform features that exhibit significant inter-individual differences. To address this issue, we further propose a phase perturbation encryptor, which functions as a phase noise generator with a learning intensity parameter to effectively conceal the potential USI embedded in the personal difference component $x_{pd}(t)$.

To extract time-varying features from the $x_{pd}(t)$, we first apply the Hilbert Transform to convert these modal signals into analytic signals on the complex plane:
\begin{equation}\label{Hilbert}
x_{pd,a} \left ( t \right ) =x_{pd} \left ( t \right ) + j\cdot \mathcal{H} \left \{ x_{pd}\left ( t \right )  \right \} ,
\end{equation}
where $x_{\text{pd,a}}(t)$ represents the analytical signal, $\mathcal{H}$ denotes the Hilbert transform operator, which is defined as follows:

\begin{equation}\label{Hilbert_1}
\mathcal{H}\left \{ x\left ( t \right )  \right \}  = \frac{1}{\pi } p.v.\int_{-\infty }^{\infty}\frac{x\left ( \tau  \right ) }{t-\tau}d\tau ,
\end{equation}
where $p.v.$ represents the Cauchy principal value integral. After obtaining the analytical signal representation $x_{pd,a}$ of $x_{pd}$, we apply the phase perturbation encryptor to inject phase noise into $x_{pd,a}$, as detailed below:

\begin{equation}\label{pert}
x_{enc,a} \left ( t \right ) = x_{pd,a}\left ( t \right ) \cdot \exp\left ( j\gamma _{f}\cdot\phi \left ( k,t \right )  \right ) ,
\end{equation}
where $\exp\big(j\gamma_{f}\cdot \phi(k,t)\big)$ denotes the phase rotation operator, with its learning intensity governed by the learnable parameter $\gamma_f$. The $\gamma_f$ is directly proportional to the identity recognition accuracy and inversely proportional to the respiration monitoring performance. Its objective is to maximize USI anonymization while minimizing the impact on key respiratory features. It is computed using the same formula as in Eq.~\ref{alpha}, but with a different threshold for the learning weight $\beta$. The $\phi(k,t)$ is the key-controlled phase perturbation function, and can be defined as follows:

\begin{equation}\label{phi}
\phi \left ( k,t \right ) = \sum_{i=0}^{L-1}k_i\cdot \frac{\pi }{2^i}\cdot g\left ( t-i\bigtriangleup t \right ) ,
\end{equation}
where $k_i$ represents the element in the key, and $g(t)$ is the Gaussian window function, which can be represented as:

\begin{equation}\label{g}
g\left ( t \right ) =\exp\left ( -\frac{t^2}{2\eta ^2_t}  \right ) ,\eta_t=\frac{T_{res}}{4}  ,
\end{equation}
where $\eta_t$ is the standard deviation of the window function, and $T_{\text{res}}$ is the duration of a complete respiratory cycle.

The perturbation is directly applied to the instantaneous phase of the signal, inducing deviations in its instantaneous frequency and thereby fundamentally altering the intrinsic features of the respiratory waveform. The instantaneous frequency deviation of the perturbed signal can be expressed as:
\begin{equation}\label{freq_change}
f_{enc}\left ( t \right ) =\frac{1}{2\pi } \frac{d}{dt} \left [ \theta \left ( t \right ) +\phi \left ( k,t \right )  \right ]=f_{inst}\left ( t \right )+\underset{\bigtriangleup f_{pert(t)}}{\underbrace{\frac{1}{2\pi }\frac{d\phi }{dt}}},      
\end{equation}
where $f_{\text{inst}}(t)$ and $\Delta f_{\text{pert}}(t)$ denote the instantaneous frequency of the raw signal and the frequency deviation induced by phase perturbation, respectively. Furthermore, the power spectrum of the perturbation satisfies the following relation:

\begin{equation}\label{offset}
P_{\bigtriangleup f}\left ( f \right ) \propto \left | \mathcal{F} \left \{ \frac{d\phi }{dt}  \right \}  \right |^2 ,
\end{equation}
where $\mathcal{F}\{\cdot\}$ denotes the Fourier transform operator.

By injecting phase noise into the universal respiratory component $x_{pd}(t)$, the overall structure of the respiratory signal is modified, thereby disrupting the original waveform patterns that could otherwise be exploited for identity recognition. To quantitatively evaluate this structural distortion, Dynamic Time Warping (DTW) is employed as follows:
\begin{equation}\label{DTW}
\mathrm {DTW}\left ( A_{enc},A \right ) =\underset{\pi}{min} \sum_{(i,j)\in \pi}\left |A_{enc}(t_i) -A(t_j) \right |^2  \ge \delta _A,
\end{equation}
where $A(t)$ and $A_{enc}(t)$ denote the original and perturbed signal envelopes, respectively, and $\pi$ represents the optimal alignment path along the time axis. Furthermore, $\delta_A$ is the lower-bound threshold for envelope similarity. In particular, a larger DTW value indicates a greater difference between the perturbed signal and its corresponding pre-perturbation signal, and the perturbed signal is accompanied by a higher degree of structural distortion, consequently signifying a stronger encryption effect.

\subsection{Irreversibility Proof}
To prevent unauthorized data acquisition, the \MB is designed to be irreversible. Even if an attacker gains access to the perturbed signal, it remains infeasible to reconstruct the raw radar data and thereby obtain the USI.

To validate the irreversibility of the perturbation, we analyze the perturbed signal from an information-theoretic perspective. Meanwhile, the mutual information is introduced as a quantitative metric to assess the dependency between the perturbed signals and its corresponding pre-perturbed signals. Specifically, since both the amplitude perturbation and phase perturbation are controlled by the encryption key $k$, the amount of original information recoverable from the perturbed signal is thus strictly bounded by the key's entropy $H(k)$. For illustrative purposes, we focus on the phase perturbation as an example to demonstrate the irreversibility of the perturbation.

Formally, the irreversibility condition can be expressed as:
\begin{equation}\label{irr_1}
I\left ( x_{enc,a};x_{pd,a} \right ) \le H\left ( k \right ) -\varepsilon ,
\end{equation}
where $I\left ( x_{enc,a};x_{pd,a} \right )$ denotes the mutual information between the phase perturbed signal $x_{enc,a}(t)$ and its corresponding unperturbed signal, \ie, the analytic signal $x_{pd,a}(t)$ of the $x_{pd}$. $H(k)$ represents the entropy of the key, and $\varepsilon$ is the security margin that defines the minimum information gap required to defend against various potential attacks.

According to \eqref{irr_1}, taking specific cases as reference, when the key length is set to $L = 128$ and the security margin is $\varepsilon = 32$ bits, the $H(k) = L \cdot \log_2 2 = 128$ bits. In this case, an attacker would need to perform at least $2^{H(k) - \varepsilon} = 2^{96}$ brute-force attempts to successfully recover the raw signal, which meets the 128-bit information security standard. Therefore, the proposed perturbation mechanism significantly achieves irreversible protection of USI, effectively mitigating the risk of privacy leakage.

\subsection{Robust Perturbation Tolerable Network (\MC)}\label{PTN_network}

After the feature decoupling and perturbation encryption, the spectral structure of the raw radar signal becomes significantly distorted, which inevitably makes conventional filter-based or phase-based processing approaches ineffective for accurate respiration detection. To achieve high-accuracy and robust respiratory monitoring under significant waveform variations introduced by encryption perturbations, by referring to the principle of Spectral Distribution Alignment, we design the \MC, which can explicitly reconstruct the spectral structure and model temporal coherence to better restore physiological consistency and extract stable respiratory features from encrypted radar signals.

As shown in Fig.~\ref{network}, the \MC contains two functional components that operate jointly to achieve distributional robustness. The first is the Spectral Distribution Alignment Block (SDAB), which performs explicit distribution modeling and alignment in the spectral domain to reshape the perturbed spectrum. The second is the Temporal Mixer Block (TMB), which enhances temporal continuity and rhythmic consistency in the reconstructed respiration waveform. These two modules are organically integrated and jointly enable the \MC to maintain signal integrity and robustness even under significant waveform variations introduced by encryption perturbations.

\subsubsection{Spectral Distribution Alignment Block (SDAB)}
To address the spectral distortion introduced by \MB, based on optimal transport theory, the first stage of \MC operates in the frequency domain is designed to explicitly model and align the signal distributions. Unlike pioneer deep networks that rely on implicit feature extraction, by performing frequency-band-wise energy alignment and redistribution, the SDAB provides an explicit and interpretable distributionally robust optimization mechanism. Specifically, the fusion signal in the time domain $\mathbf{X}$ is first transformed into the time–frequency domain through the short-time Fourier transform (STFT) to obtain the amplitude spectrum $\mathbf{M}$ and phase spectrum $\mathbf{\Phi}$. The average amplitude spectrum $\bar{\mathbf{M}}$ is then computed along the temporal dimension. In order to better conceal the USI, considering that respiratory information resides within a fixed frequency band, the spectrum is divided into two bands: the respiration band $F_b$ and the other band $F_o$. Thus, their normalized energy distributions can be expressed as:
\begin{align}
\mathbf{p}_{b} = \frac{\bar{\mathbf{M}}[:, :, F_b]}{\sum{f \in F_b} \bar{\mathbf{M}}[:, :, f]}, \\
\mathbf{p}_{o} = \frac{\bar{\mathbf{M}}[:, :, F_o]}{\sum{f \in F_o} \bar{\mathbf{M}}[:, :, f]}.
\end{align}

Then, based on the two different frequency bands, we define different target distributions for them, \ie, $\mathbf{q}_{b}$ and $\mathbf{q}_{o}$, which are used to optimize the corresponding components, \ie, the respiratory component $x_{re}(t)$ in the respiratory frequency band, and the other unrelated components $x_{ot}(t)$ outside the respiratory frequency band, guiding them toward the ideal distribution. For the respiration band spectrum $F_b$, we introduce a learnable prototype distribution $\mathbf{q}_b = \text{Softmax}(\mathbf{\theta}_{b})$ to preserve the morphology of respiratory patterns, where $\mathbf{\theta}_{b}$ denotes trainable parameters. In contrast, for the other band spectrum $F_o$, a uniform distribution $\mathbf{q}_o \in \mathbb{R}^{|F_o|}$ is employed to suppress person-specific spectral variations and reduce identifiability. To better optimize these distributions, the entropy-regularized Sinkhorn Optimal Transport formulation is used due to more stable in the face of perturbations, sampling noise or incomplete distribution matching. Specifically, the spectral alignment cost matrix is defined as $C[j,k] = (f_j - f_k)^2$, and the regularized optimal transport objective $W_{\nu}(\mathbf{p}, \mathbf{q})$ can be written as:
\begin{equation}\label{w}
W_{\nu}(\mathbf{p}, \mathbf{q}) =
\min_{\xi \in \Pi(\mathbf{p}, \mathbf{q})}
\langle \xi, C \rangle + \nu H(\xi),
\end{equation}
where $H(\xi)$ is the entropy term and $\nu$ is a parameter that balances smoothness and alignment precision. 


To enhance the dominant peaks in the respiration band while suppressing identity cues outside the respiration band, thus facilitating the extraction of stable respiratory features from encrypted radar signals, based on the aligned distributions, the SDAB generates a spectral reallocation mask $\mathbf{M}_{ad}(f)$ to adjust the energy ratio of each frequency component to better reconstruct the spectral structure, which can be expressed as:



\begin{equation}
\mathbf{M}_{ad}(f) =
\begin{cases}
\text{clip}\left(\frac{\mathbf{q}_b}{\mathbf{p}_b}, [0.5, 2.0]\right), & f \in F_b;\\
\text{clip}\left(\frac{\mathbf{q}_o}{\mathbf{p}_o}, [0.25, 1.0]\right), & f \in F_o.
\end{cases}
\end{equation}
where the clip operation constrains the values of the spectral reallocation mask within predefined bounds to ensure a stable reconstruction of the spectral structure.

To obtain the adjusted amplitude spectrum $\mathbf{M}_{al}$, the original amplitude spectrum $\mathbf{M}$ is combined with the spectral reallocation mask $\mathbf{M}_{ad}(f)$ through the Hadamard product \cite{10965485}, \ie, $\mathbf{M}_{al} = \mathbf{M} \odot \mathbf{M}_{ad}(f)$. Then, to reconstruct the complete complex spectrum, this adjusted spectrum is then combined with the original phase $\mathbf{\Phi}$ in its polar form \cite{lu2025explicit}, denoted as:
\begin{equation}
\mathbf{X}_{\text{rec}} = \text{Polar}(\mathbf{M}{al}, \mathbf{\Phi}).
\end{equation}

Finally, by transforming the $\mathbf{X}_{\text{rec}}$ back to the time domain through the inverse STFT, the reconstructed signal $\hat{\mathbf{x}}$ is obtained. This explicit spectral alignment and energy redistribution reinforce the respiratory features in the respiratory frequency band while suppressing other features outside the respiratory frequency band.

\subsubsection{Temporal Mixer Block (TMB)}

Although the SDAB restores the overall spectral structure, the reconstructed waveform may still exhibit slight temporal discontinuities and amplitude drift. To refine temporal consistency, a TMB is introduced. This module is composed of several layers of one-dimensional convolution, gated linear units (GLU), and batch normalization (BN), combined with residual connections to maintain stability. The input feature sequence first undergoes convolution to capture local temporal context, then passes through the GLU to adaptively modulate temporal dependencies, followed by BN and residual summation, which can be represented as:

\begin{equation}
\mathbf{Y} = \text{BN}\big(\text{GLU}(\text{Conv1D}(\hat{\mathbf{x}}))\big) + \hat{\mathbf{x}}.
\end{equation}

\begin{table}[t]
\renewcommand{\arraystretch}{1.8}
\caption{Configuration of mmWave radar.}
\centering
\begin{tabular}{cccc} 
\toprule[1.5pt]
Parameters & Values & Parameters   & Values  \\ 
\midrule[1.5pt]
Starting Frequency & $77Ghz$   &   BandWidth  &  $4Ghz$  \\ 
\midrule
RX Gain   & $50dB$  &  Idle Time  &  $10\mu s$    \\ 
\midrule
Chirp Cycle Time  &  $50\mu s$   &  Chirps/Frame  &  $128$  \\
\midrule
Frame Periodicity  &  $50ms$   &   Samples/Chirp   &   256   \\
\midrule
ADC Sample rate  &  $8000K$   &   Frequency slop  &  $80MHz/\mu s$\\
\bottomrule
\end{tabular}
\label{configure}
\end{table}

This structure captures the periodic and rhythmic characteristics of respiration, improves temporal smoothness, and enhances robustness against perturbations.

In summary, the \MC integrates SDAB and TMB to comprehensively achieve explicit spectral reconstruction and enhance temporal robustness. The SDAB performs Optimal Transport-based alignment and energy redistribution in the spectral domain, while the TMB strengthens coherence in the time domain. Furthermore, they jointly constitute a distributionally robust and physically interpretable perturbation-tolerant mechanism that ensures respiratory features remain recoverable and stable under privacy-preserving perturbations.

\begin{figure*}[t]
	\centering
	\subfloat[Identity recognition performance of baseline.]
        {\label{baseline_con}
        \includegraphics[width=0.32\textwidth]{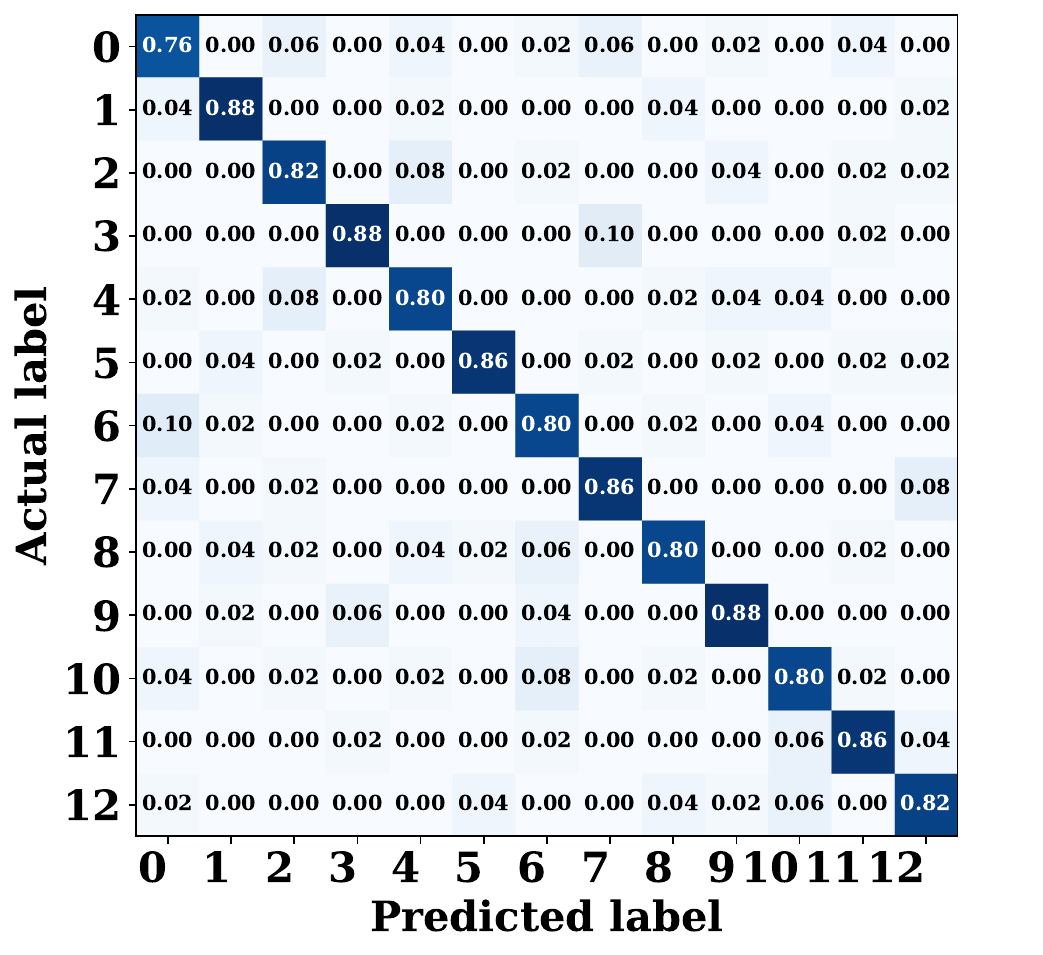}}
	\subfloat[Identity recognition performance of \name.]
        {\label{name_con}
        \includegraphics[width=0.32\textwidth]{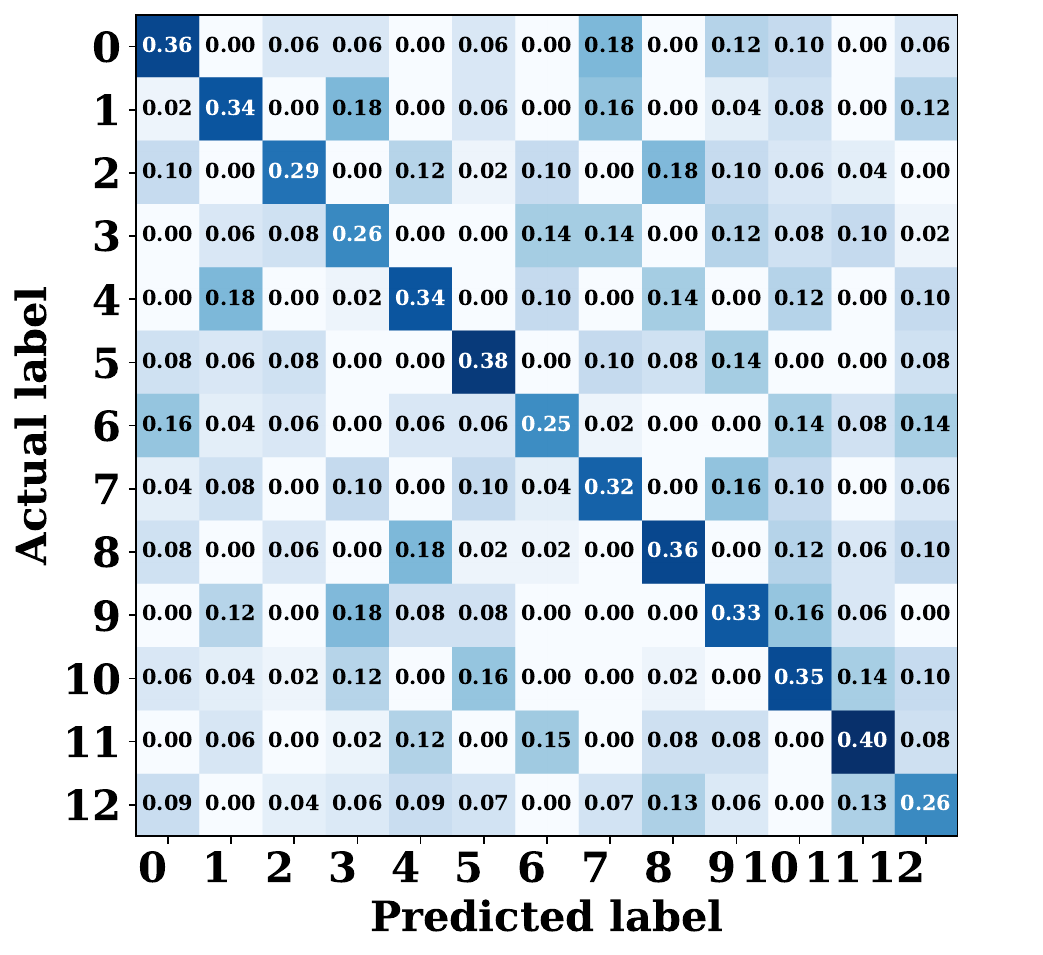}}
    \subfloat[MAE performance of \name.]
        {\label{CDF_overall}
        \includegraphics[width=0.3\textwidth]{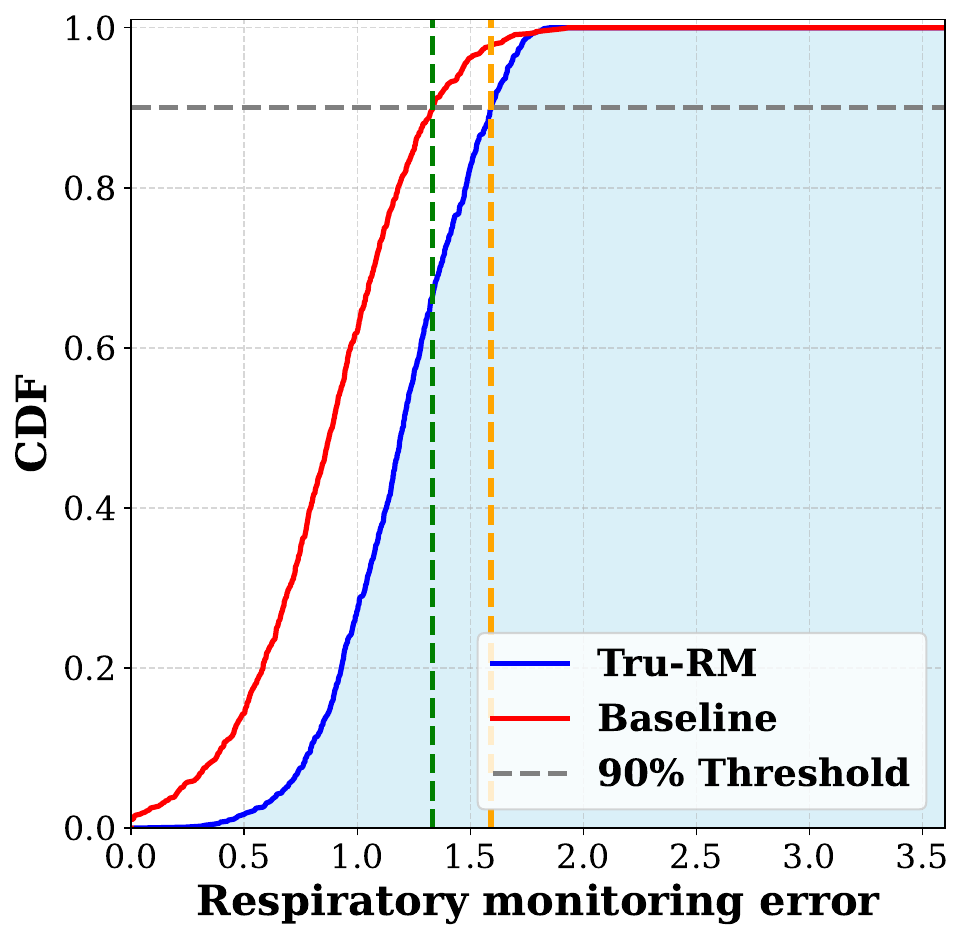}}
	\caption{Overall performance comparison with baseline in privacy protection and respiratory monitoring.}
    \label{confusion_matrix}
\end{figure*}

\subsubsection{Identity Recognition}
For a fair comparison of identity concealment before and after encryption, we adopt a one-dimensional convolutional neural network for identity recognition. The perturbed radar signals are processed through multiple convolutional and pooling layers to progressively extract the key features and capture identity-related representations. The network consists of three convolutional blocks, each comprising convolution, normalization, and ReLU activation operations to enhance discriminative capability. The extracted features are then aggregated by an adaptive average pooling layer and subsequently passed through two fully connected layers for feature mapping and classification.

\subsubsection{Loss Function}
The proposed \name performs respiratory monitoring and identity recognition based on the perturbed signals, and then updates the parameters of the learnable noise intensity and \MC dynamically according to the training feedback. The overall process is supervised by four losses: identity loss, respiratory loss, spectral distribution alignment loss, and waveform loss.

\textbf{\emph{Identity Loss:}}
Identity loss is used to measure the similarity between the predicted identities and the true labels, which is quantified using the cross-entropy loss:
\begin{equation}\label{id loss}
\begin{aligned}
\mathcal{L}_{\text{id}}=-\frac{1}{N}\sum_{i=1}^{N}logp_i(y_i)  ,
\end{aligned}
\end{equation}
where $N$ represents the batch size, $p_i$ and $y_i$ denote the predicted result and true identity label.

\textbf{\emph{Respiratory Loss:}}
To measure the accuracy of respiratory rate estimation, the Mean Absolute Error (MAE) is employed as the primary regression objective:
\begin{equation}\label{freq loss}
\mathcal{L}_{\text{r}}=\frac{1}{N}\sum_{i=1}^{N}|\hat{f}_i - f_i|,
\end{equation}
where $\hat{f}i$ and $f_i$ denote the predicted rate and the true label.

\textbf{\emph{Spectral Distribution Alignment Loss:}}
The spectral distribution alignment loss guides the network to enhance respiratory-band energy while suppressing identity-related spectral components, thus ensuring distributional robustness against perturbations. The alignment loss is formulated as:
\begin{equation}\label{sda loss}
\mathcal{L}_{\text{SDA}} =
W{\nu}(\mathbf{p}_{b}, \mathbf{q}_{b}) +
W_{\nu}(\mathbf{p}_{i}, \mathbf{q}_{i}).
\end{equation}
where $W{\nu}(\mathbf{p}, \mathbf{q})$ denotes the regularized optimal transport objective defined in Eq.~\ref{w}.

\begin{table}[t]
\renewcommand{\arraystretch}{1.8}
\caption{Performance comparison of different encryption algorithm for privacy protection.}\label{tab:MB}
\centering
\begin{tabular*}{0.9\linewidth}{@{\extracolsep{\fill}} c c c c c}
\toprule
Method & IRAC  & \makecell{Encryption \\Time (ms)} & MAE (bpm) & STD (bpm)  \\ 
\midrule
w/o s1+s2 &  $83.38\%$  & 0.00 & $0.9$ &   $0.36$  \\ 
\midrule
w/o s1 &  $47.97\%$  &  19.28 &  $1.14$ &   $0.31$  \\ 
\midrule
w/o s2 &  $76.24\%$  &  30.09 &  $0.96$ &   $0.34$   \\ 
\midrule
\textbf{s1+s2 (\MB)} &  $\textbf{32.62\%}$  &  \textbf{51.42}  &  $\textbf{1.2}$ &   $\textbf{0.3}$ \\ 
\midrule
MuFilter\cite{liu2025multi} &  $27.37\%$  & 24.16  &  $\geq 3$  &   $\geq1$  \\ 
\midrule
mmFilter\cite{liu2023application} &  $27.29\%$  & 23.95  &  $\geq 3$  &   $\geq1$  \\ 
\bottomrule
\end{tabular*}
\label{overall perform}
\end{table}


\textbf{\emph{Waveform Loss:}} 
The waveform loss is used to quantify the similarity between the perturbed signal waveform and the original radar signal waveform. The loss is formulated as:
\begin{equation}\label{morph loss}
\begin{aligned}
\mathcal{L}_{mor}=\underset{\pi \in \mathcal{A} }{min} \sum_{(i,j)\in \pi}\left \| S(t_i)-S_{enc}(t_j) \right \|^2+ \\ \lambda_w \underset{\rho \in\tau }{inf} \mathbb{E}_{(p,q)\sim \rho } \left [ \left \| \bigtriangledown _t\mathcal{N}-\bigtriangledown _t\mathcal{N}_{enc} (q) \right \|  \right ]  ,
\end{aligned}
\end{equation}
where $\mathcal{A}$ is the set of DTW alignment paths, $S(t)$ and $S_{enc}(t)$ represent the signal envelopes of the original and perturbed radar waveforms, respectively. $\tau$ denotes the set of joint distributions, $\bigtriangledown _t\mathcal{N} (q)$ and $\bigtriangledown _t\mathcal{N}_{enc} (q)$ represent the temporal gradients of the original and perturbed radar signals, respectively.

\textbf{\emph{Overall Objective:}}
Overall, the objective for the \name is formulated as follows:
\begin{equation}\label{loss}
\mathcal{L}_{total}=\lambda_{id}\mathcal{L}_{id}+\lambda_r\mathcal{L}_{r}+\lambda_s\mathcal{L}_{SDA}+\lambda_{mor}\mathcal{L}_{mor},
\end{equation}
where $\lambda(\ast)$ is the hyperparameter which controls the magnitude of each loss. The overall training objective is to maximize the identity recognition error while minimizing the respiratory monitoring error.

\begin{table}[t]
\renewcommand{\arraystretch}{1.8}
\caption{Performance comparison with baseline in privacy protection and respiratory monitoring.}
\centering
\begin{tabular*}{0.9\linewidth}{@{\extracolsep{\fill}} c c c c}
\toprule
Method & IRAC  & MAE (bpm) & STD (bpm)  \\ 
\midrule
Baseline & $83.38\%$   &    $0.9$ &   $0.36$  \\ 
\midrule
\textbf{\name} &  $\textbf{32.62\%}$  &   $\textbf{1.2}$ &   $\textbf{0.3}$  \\ 
\midrule
Improvement &  $50.76\%\uparrow$  &  $0.3 \downarrow$    &  $0.06\uparrow$  \\
\bottomrule
\end{tabular*}
\label{overall perform}
\end{table}

\begin{figure}[t]
    \vspace{-0.2in}
	\centering
    \subfloat[Performance comparison of encryption algorithm.]
        {\label{ablation}
        \includegraphics[width=0.24\textwidth]{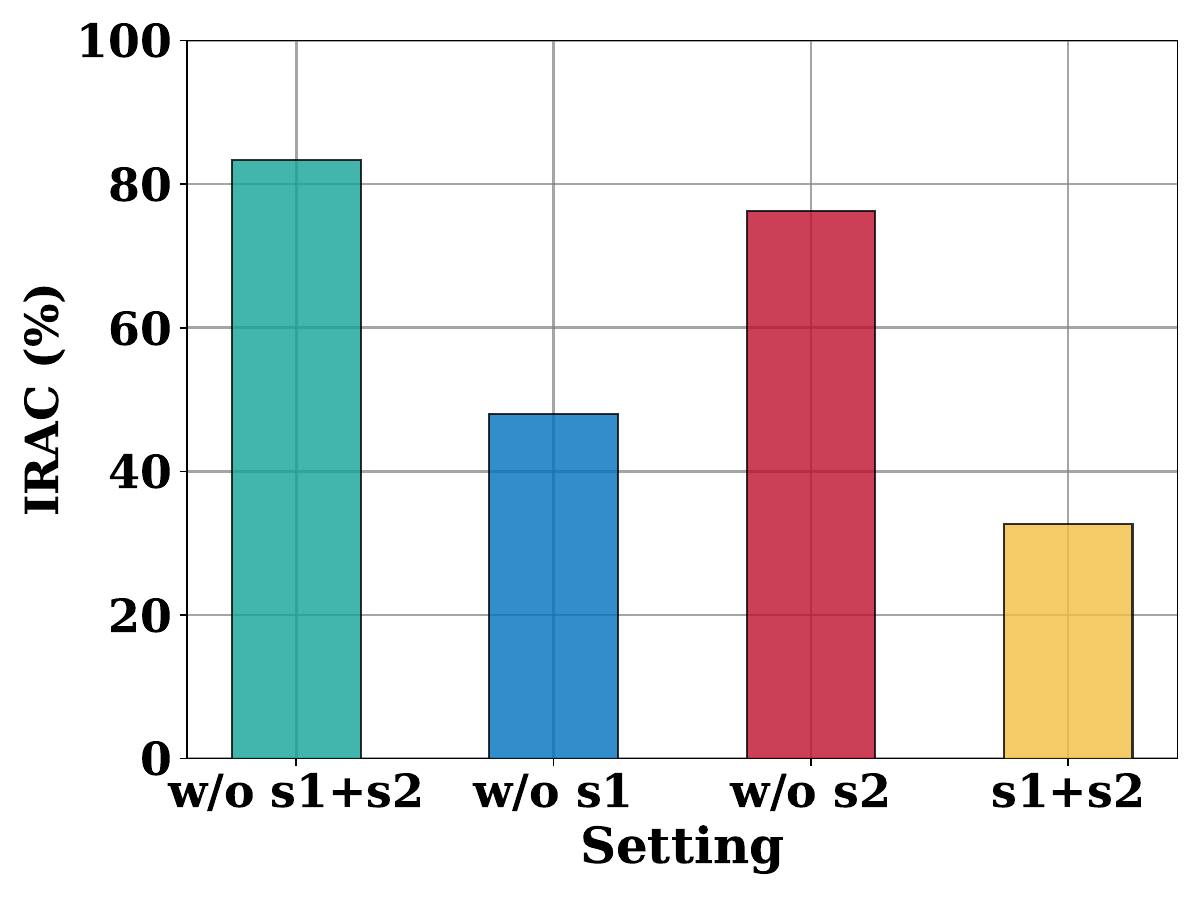}}
    \subfloat[Performance comparison of feature extraction algorithm.]
        {\label{ablation_PTN}
        \includegraphics[width=0.235\textwidth]{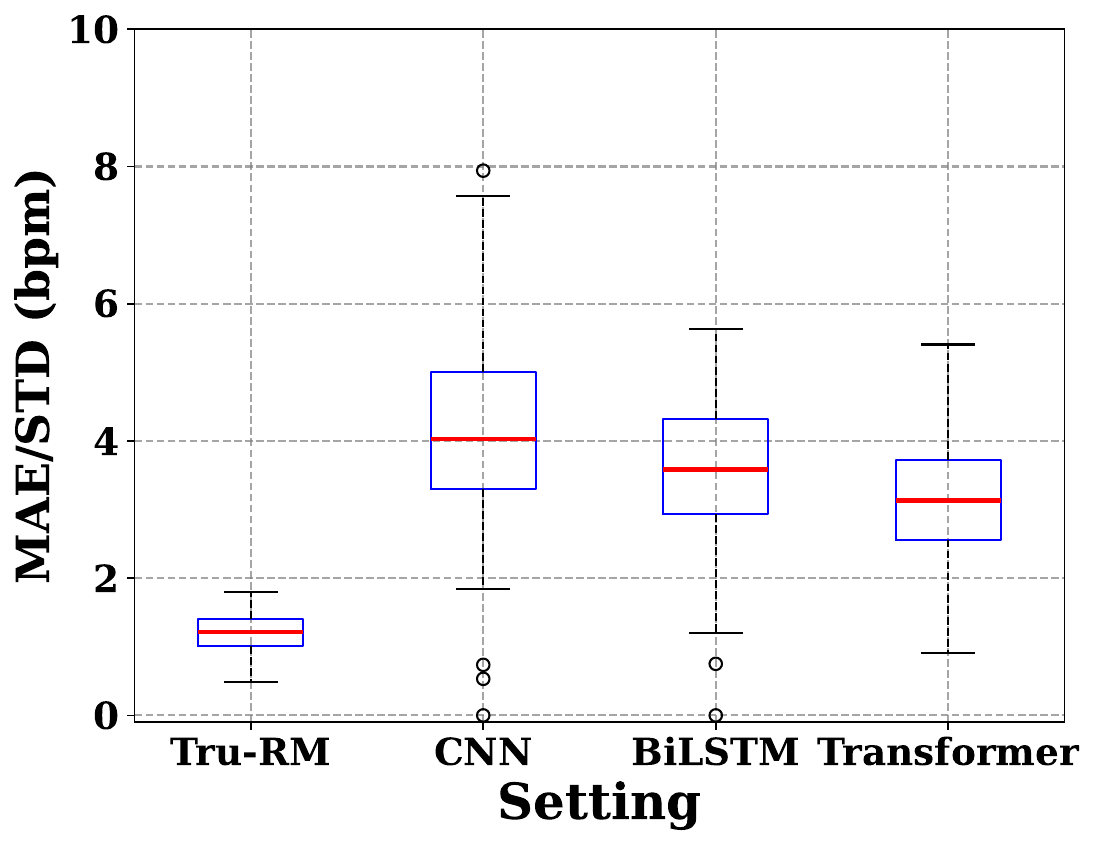}}
	\caption{Performance comparison of encryption algorithm and respiratory feature extraction algorithm for perturbed signals.}
\end{figure}

\section{Performance Evaluation}\label{Experimental Evaluation}

In this section, we first evaluate the overall performance of \name, including the effectiveness of anonymization and the performance of the respiratory monitoring under significant waveform variations. Then, we validate the effectiveness of the key modules within \name. Finally, to further investigate the system's robustness across different measurement distances, respiratory patterns, and monitoring durations, we conduct extensive experiments to comprehensively assess its performance.

\subsection{Experimental Setup}
To implement \name, we employ a commodity off-the-shelf (COTS) mmWave radar, \ie, the TI AWR1843 BOOST, which integrates three transmitting antennas (TX) and four receiving antennas (RX). The detailed radar configuration parameters are provided in Tab.~\ref{configure}. For data acquisition, the DCA1000EVM board is used to capture the raw mmWave signals. Then, we process these raw radar signals on a desktop PC equipped with an Intel i9-9900KF CPU and 24 GB of RAM.

A total of 13 individuals participate in the study, including 4 females and 9 males, with body weights ranging from 40 kg to 80 kg and ages between 18 and 28 years old. Specifically, all subjects have signed informed consent forms before the start of the experiment. During the experiment, each participant is instructed to sit on a chair in a relaxed posture, avoiding any large movements to avoid motion interference. The mmWave radar is placed on a desk directly in front of the participant, with the antenna facing the chest. Furthermore, a Bigrun Team device is adopted as the ground truth, which is worn on the subject’s chest to obtain accurate respiratory measurements. In each data collection session, the participant is instructed to remain still for 2-15 minutes, followed by a 10-minute rest between consecutive data collection sessions. Data collection is carried out across different days and at varying times, yielding a total of 30 minutes of data per participant. Unless mentioned otherwise, we split the dataset by a 8/2 ratio, \ie, $80\%$ of the data are used to construct models, and $20\%$ are employed to evaluate the performance.

\begin{figure}[t]
\vspace{0.1in}
	\centering
    \subfloat[Identity recognition performance.]
        {\label{duration_zhu}
        \includegraphics[width=0.245\textwidth]{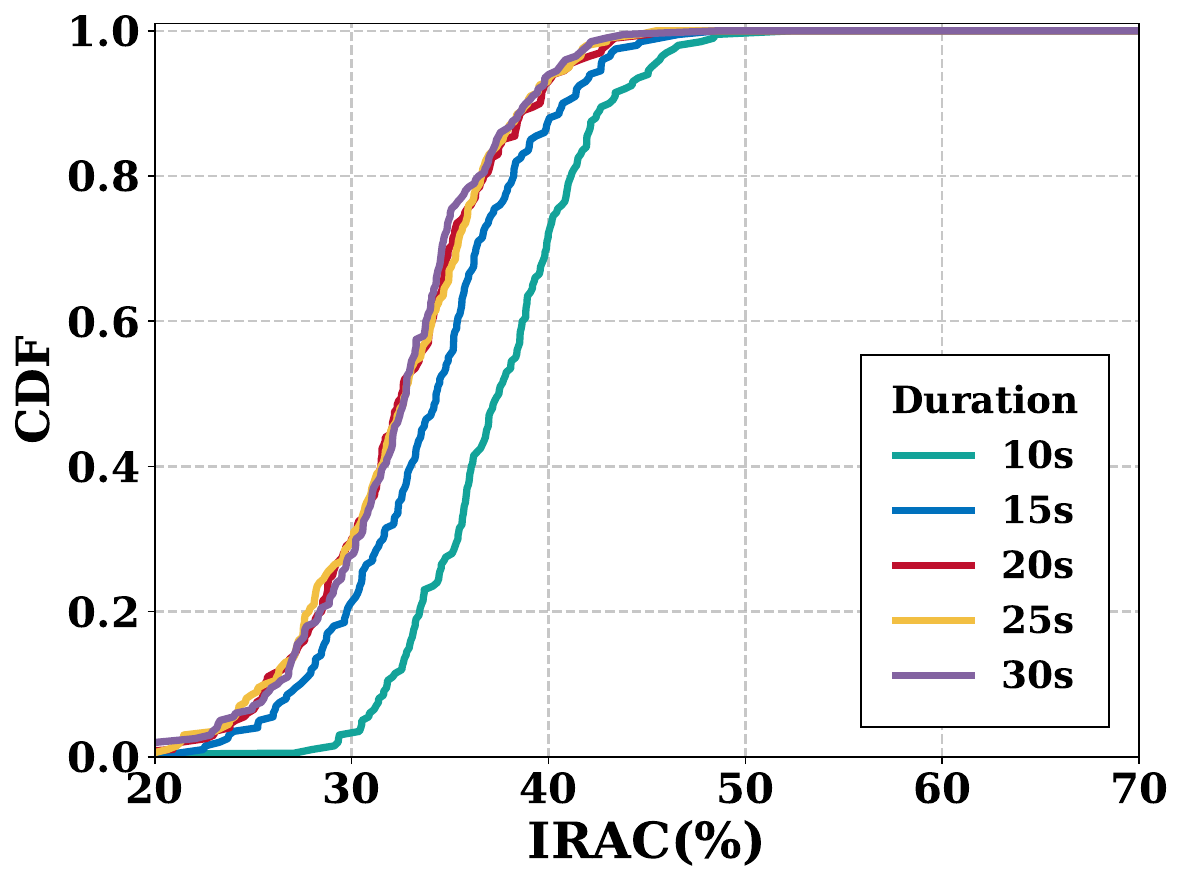}}
    \subfloat[Respiratory monitoring performance.]
        {\label{duration_box}
        \includegraphics[width=0.25\textwidth]{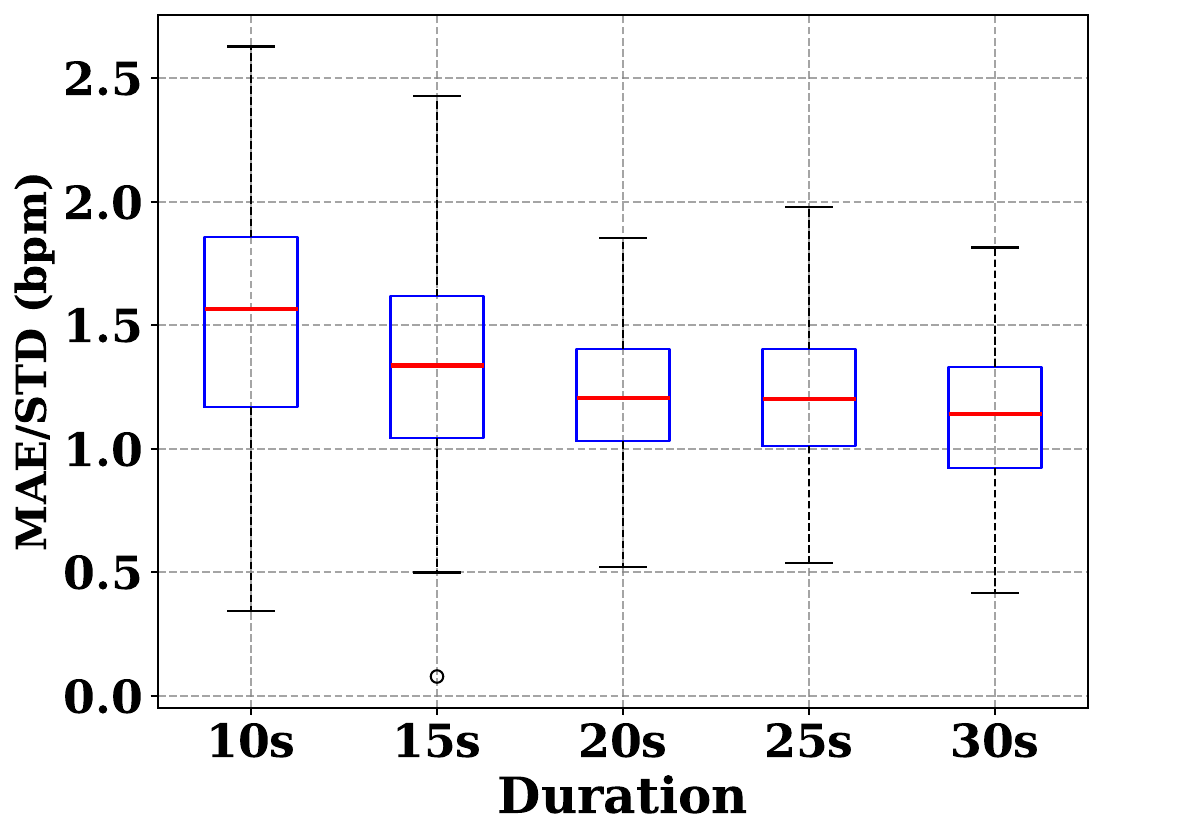}}
	\caption{Performance comparison on different durations.}
    \label{duration}
\end{figure}

\begin{table}[t]
\renewcommand{\arraystretch}{1.8}
    \vspace{0.2in}
    \centering
    \captionof{table}{Performance comparison of respiratory feature extraction algorithm for perturbed signals.}
    \label{tab:MC}
    \vspace{-0.1cm}
        \begin{tabular*}{0.9\linewidth}{@{\extracolsep{\fill}}ccccc}
        \toprule
        \multirow{2}{*}{} & \multicolumn{4}{c}{Respiratory monitoring model} \\
        \cmidrule(l){2-5} 
          & \textbf{\name}  & CNN &  BiLSTM  &  Transformer  \\ \midrule
        MAE (bpm) & \textbf{1.20} & 4.26 & 3.60 & 3.18 \\  
        STD (bpm) & \textbf{0.30} & 1.20 & 0.96 & 0.86 \\  
        Time (ms)  & \textbf{86.88} & 67.28 & 65.32 & 78.54\\
        \bottomrule
        \end{tabular*}
\end{table}

\subsection{Metrics}
To evaluate the respiratory monitoring performance, we adopt two different metrics, \ie, the Mean Error (MAE) $\mu =\frac{ {\textstyle \sum_{j=1}^{J}|(\widehat{b_j} -b_j)|} }{J}$ and the Standard Deviation of mean error (STD) $\sigma =\sqrt{\frac{ {\textstyle \sum_{j=1}^{J}(\widehat{b_j} -b_j-\mu)^2} }{J} } $. The $\widehat{b_j}$ represents the predicted value of the respiratory rate, $b_j$ represents the corresponding ground truth, and $J$ is the total number of samples. Both MAE and STD are measured in beats per minute (bpm).

The identity recognition is a multi-classification task, which aims to evaluate the model's ability to recognize identity information. In this task, we adopt the identity recognition accuracy (IRAC) as the evaluation metric, which can be denoted as:

\begin{equation}
IRAC=\frac{1}{N}\sum_{i=1}^{N} \mathbb{I}\left ( \hat{c_i}=c_i  \right )   , 
\end{equation}
where $c_i$ and $\hat{c}_i$ denote the true label and the predicted label of the $i$-th sample, respectively, and $\mathbb{I}(\cdot)$ is the indicator function. It is worth noting that, in this study, a higher IRAC value is not necessarily desirable. Instead, it serves as a reverse measurement indicator of the privacy protection effect. The lower the IRAC, the more difficult the user identity information can be recognized, and the better the performance of the model in terms of privacy protection.

\begin{figure}[t]
\vspace{0.1in}
	\centering
    \subfloat[Identity recognition performance.]
        {\label{distance_zhu}
        \includegraphics[width=0.245\textwidth]{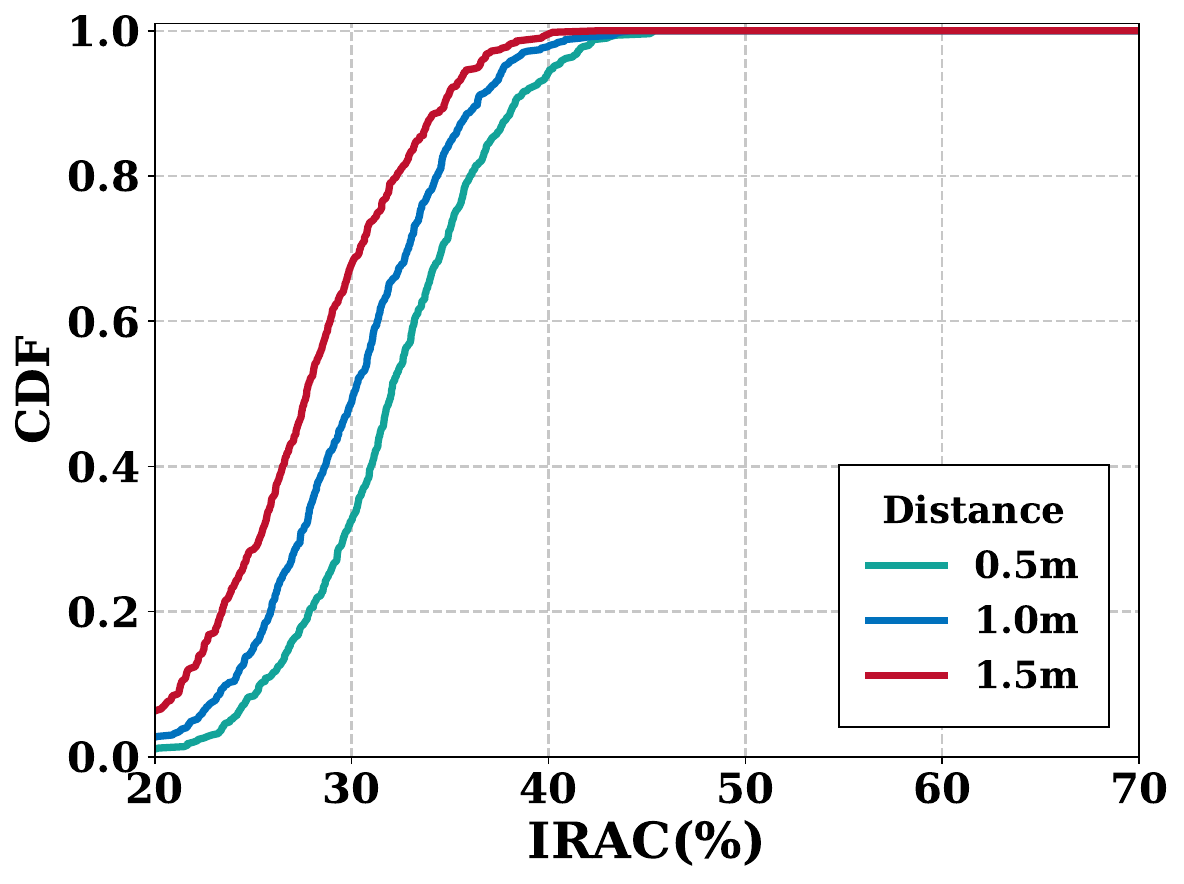}}
    \subfloat[Respiratory monitoring performance.]
        {\label{distance_line}
        \includegraphics[width=0.255\textwidth]{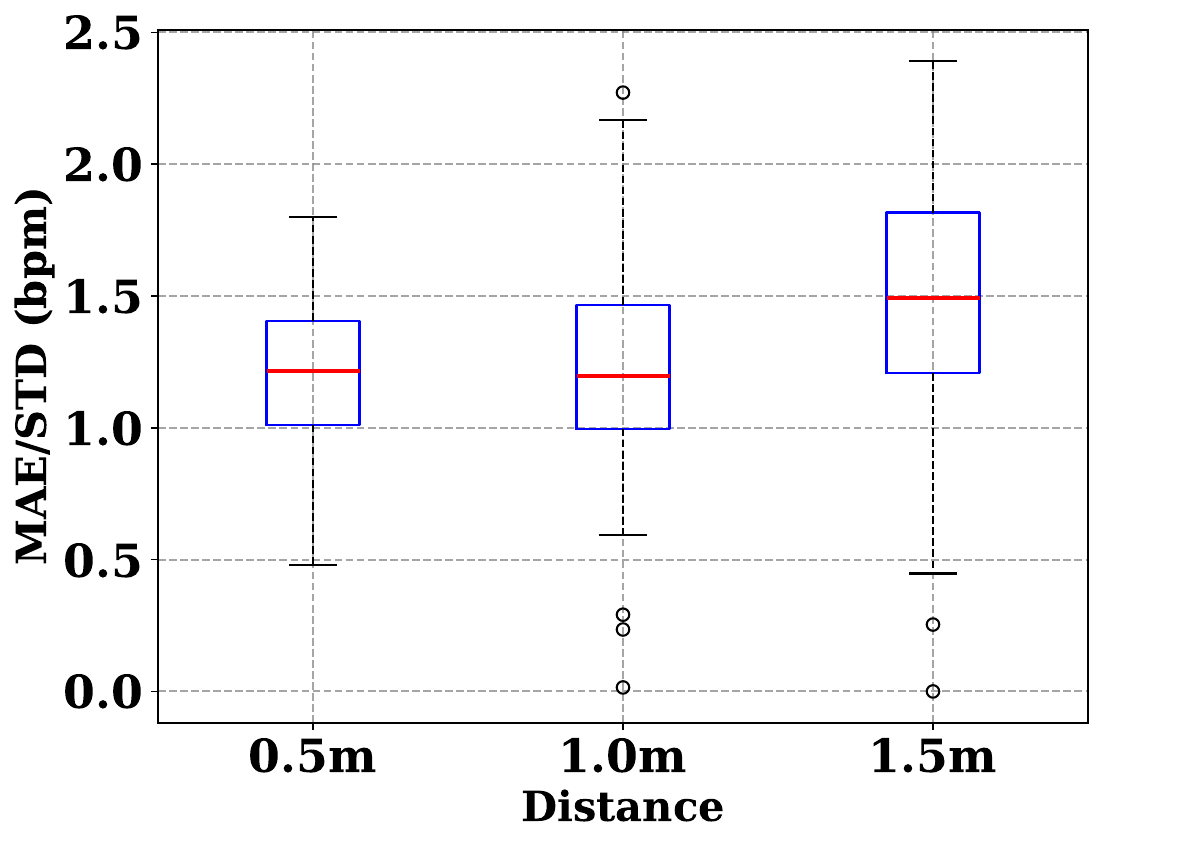}}
	\caption{Performance comparison on different distances.}
    \label{distance}
\end{figure}

\begin{figure}[t]
\vspace{0.1in}
	\centering
    \subfloat[Identity recognition performance.]
        {\label{pattern_zhu}
        \includegraphics[width=0.25\textwidth]{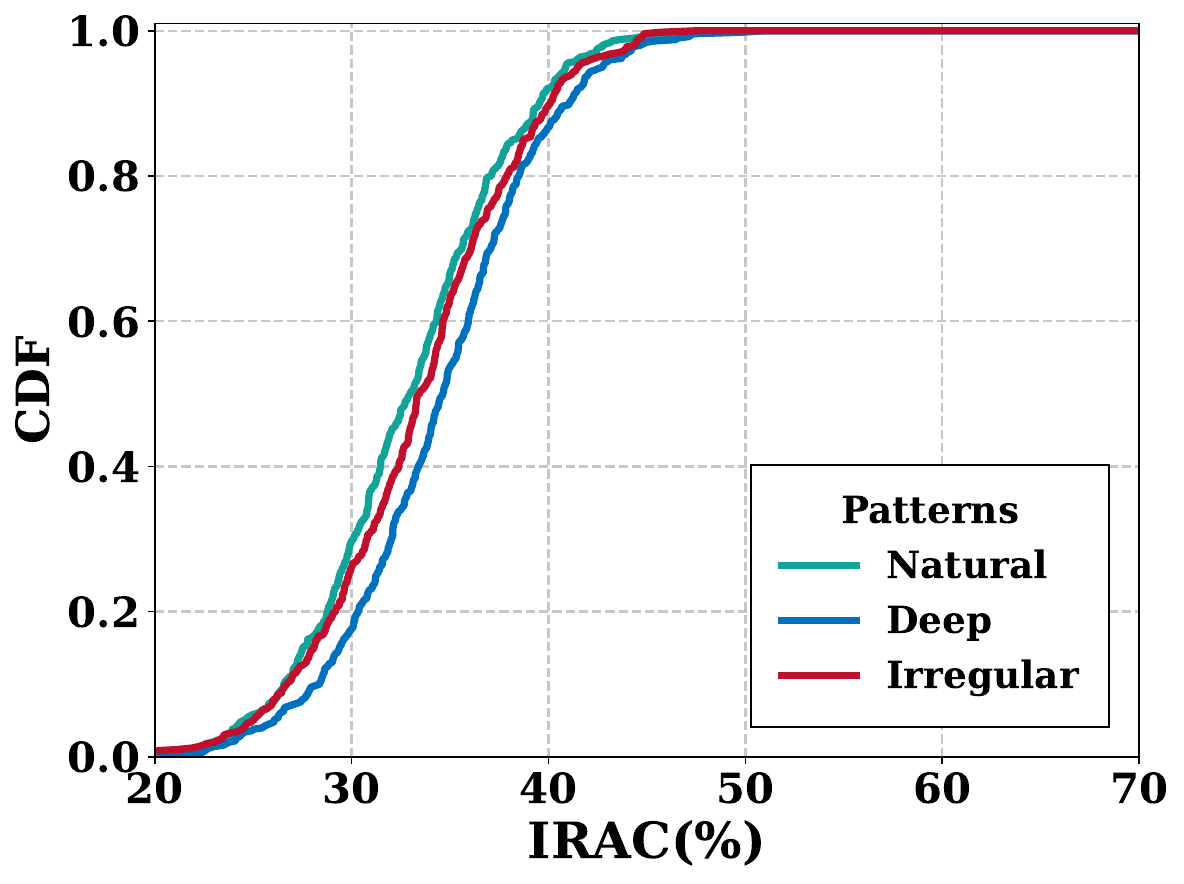}}
    \subfloat[Respiratory monitoring performance.]
        {\label{pattern_box}
        \includegraphics[width=0.25\textwidth]{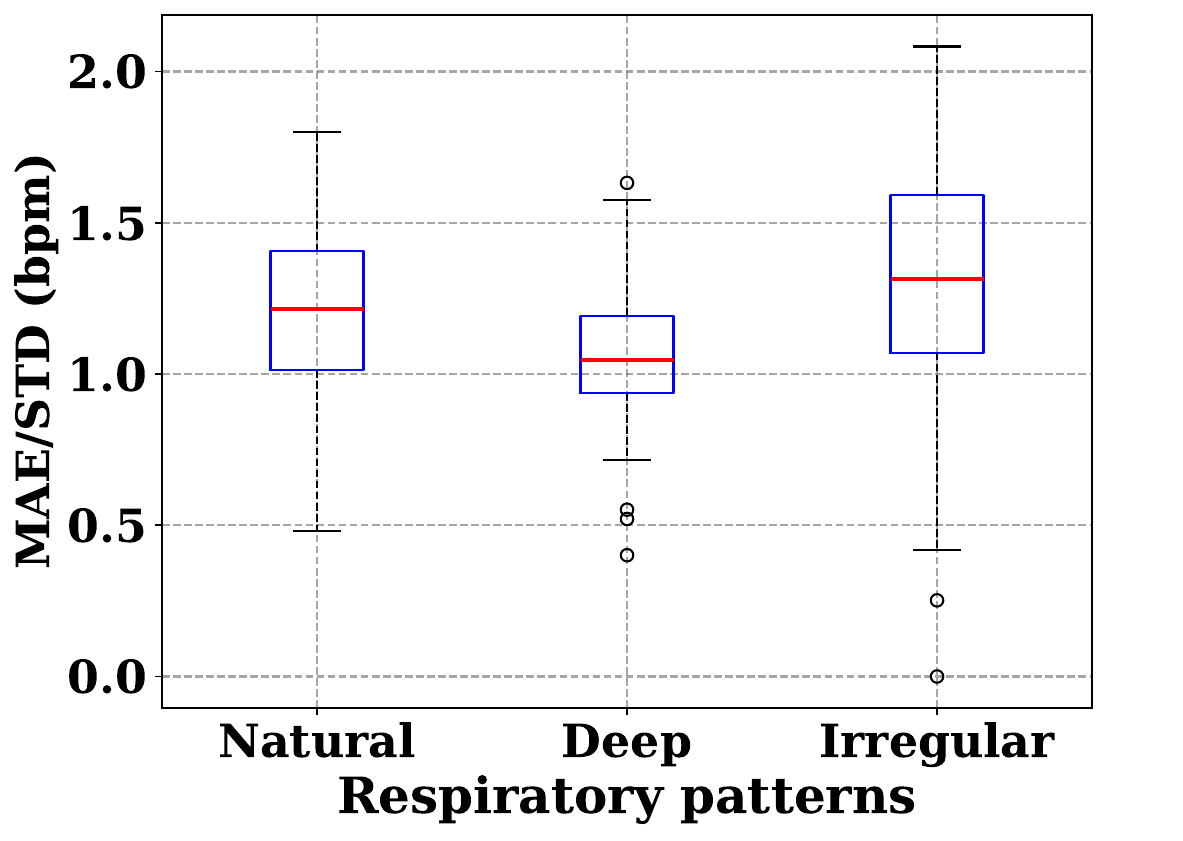}}
	\caption{Performance of comparison on different respiratory patterns.}
    \label{pattern}
    \vspace{-0.2in}
\end{figure}

\subsection{Overall Performance} 
To quantitatively evaluate the overall performance of \name, we compare it against baseline where the raw radar data is directly used for identity recognition and respiratory monitoring without any perturbations.

As shown in Tab.~\ref{overall perform}, the \name strikes a favorable balance between user privacy protection and respiratory monitoring. Specifically, while the IRAC value drops sharply from a baseline of $83.38\%$ to only $32.62\%$ after encryption, the respiratory monitoring performance remains largely unaffected. In particular, the MAE of respiratory monitoring is $1.2bpm$, almost identical to the baseline. Moreover, the \name yields lower STD values, which clearly highlights its ability to maintain robust and reliable respiratory estimation. Apparently, this performance benefits from the designs of both \MB and \MC, \ie, the former effectively preserves key respiratory features by injecting targeted encryption, while the latter module, which is grounded in spectral distribution alignment principle to restore physiological consistency, effectively enables high-accuracy respiratory monitoring even under significant waveform variations introduced by perturbations.

\begin{table*}[t]
\renewcommand{\arraystretch}{1.8}
    \centering
    \captionof{table}{Performance comparison on different measurement distances, respiratory patterns and durations.}
    \label{tab:different}
    \vspace{-0.1cm}
        \begin{tabular*}{0.99\linewidth}{@{\extracolsep{\fill}}cccc|ccc|ccccc}
        \toprule
        \multirow{2}{*}{} & \multicolumn{3}{c}{Measurement Distance} & \multicolumn{3}{c}{Respiratory pattern} & \multicolumn{5}{c}{Duration} \\
        \cmidrule(l){2-12} 
          & $0.5m$ & $1.0m$ & $1.5m$ & Natural & Deep  & Irregular &  $10s$  &  $15s$  &  $20s$   & $25s$ & $30s$ \\ \midrule
        MAE (bpm) & 1.20 & 1.26 & 1.50 & 1.20 & 1.08 & 1.32 & 1.50 & 1.32 & 1.20 & 1.20 & 1.14 \\  
        STD (bpm) & 0.30 & 0.33 & 0.42 & 0.30 & 0.18 & 0.36 & 0.48 & 0.42 & 0.30 & 0.30 & 0.30\\  
        IRAC & $32.62\%$ & $30.14\%$ & $27.97\%$ & $32.62\%$ & $34.91\%$  & $33.71\%$ & $37.75\%$ & $34.02\%$ & $32.97\%$ & $32.71\%$ & $32.62\%$\\
        \bottomrule
        \end{tabular*}
\end{table*}

To more intuitively demonstrate the performance of \name, we conduct a visual analysis of the system performance. Fig.~\ref{baseline_con} and Fig.~\ref{name_con} show the confusion matrices of identity recognition results before and after applying the proposed encryption process, \ie, \MB, respectively. The horizontal and vertical axes correspond to the predicted identity labels and the true identity labels. Compared with the non-encrypted baseline, the diagonal dominance in the confusion matrix of \name is substantially weakened, and the IRAC is markedly reduced, which means that the identity information in the radar signals has been effectively concealed. This clearly indicates that the \MB can effectively suppress USI. Next, to intuitively compare the performance of respiratory monitoring, we plot the empirical Cumulative Distribution Function (CDF) of the respiratory monitoring error, as shown in Fig.~\ref{CDF_overall}. The horizontal axis represents the respiratory monitoring error, while the vertical axis denotes the cumulative probability, indicating the proportion of samples with errors below a given value. It is evident that although the \name exhibits higher respiratory monitoring errors compared with the baseline, $90\%$ of the errors remain below $1.6 bpm$, demonstrating that \MB effectively preserves key respiratory features during perturbation, while \MC enables high-accuracy respiratory monitoring even from noised signals.

\subsection{Effectiveness of Key Module in \emph{\name}}
\subsubsection{Effectiveness of FPE}
As described in Sec.~\ref{encryptor}, we propose the \MB to perturb the raw radar signals jointly from the amplitude and phase domains to conceal USI. To evaluate the contribution of each key encryption component in \MB, we design four experimental configurations, \ie, no encryption (w/o s1+s2), encryption without the amplitude perturbation (w/o s1), encryption without the phase perturbation (w/o s2), and the \name (s1+s2). Fig.~\ref{ablation} and Tab.~\ref{tab:MB} illustrate the comparison of encryption performance under these four settings. In Fig.~\ref{ablation}, the horizontal axis denotes the different settings, while the vertical axis represents the corresponding IRAC values. The results indicate that, without encryption, the model achieves an IRAC of $83.38\%$, suggesting that the raw radar signals contain abundant USI that can be used for identity recognition. When amplitude or phase perturbations are applied separately to $x_{ot}$ and $x_{ure}$, the IRAC dropped to $76.24\%$ and $47.97\%$, respectively, indicating that both $x_{ot}$ and $x_{ure}$ contain USI, and the amplitude and phase perturbations can effectively conceal USI in the corresponding domain. Furthermore, when the perturbation acts on both the amplitude and phase domains simultaneously, the IRAC significantly drops to $32.62\%$, demonstrating that \MB can effectively conceal USI in both the phase and amplitude domains, achieving the highest anonymization performance and thereby validating the effectiveness of the \MB.

To further validate the effectiveness of the proposed perturbation encryption method in preserving respiratory features, we compare the respiratory monitoring performance of radar signals encrypted by \MB with that of state-of-the-art privacy protection approache, \eg,  MuFilter \cite{liu2025multi} and mmFilter\cite{liu2023application} based on phase fluctuation, as illustrated in Tab.~\ref{tab:MB}. The experimental results indicate that, even after \MB encryption, the radar signals still achieve high-accuracy and robust respiratory monitoring. In contrast, the MuFilter and mmFilter almost completely destroy the radar signal characteristics, resulting in a severe degradation of respiratory monitoring performance. Specifically, compared to the MuFilter and mmFilter, \MB does not significantly increase the encryption time. The result clearly demonstrates that when \MB introduces targeted perturbations primarily into the components containing USI while minimally affecting respiration-related features, thus achieving both effective and lightweight USI anonymization and high-accuracy respiratory monitoring.

\subsubsection{Effectiveness of \MC}
As described in Sec.~\ref{network}, to achieve high-accuracy trusted respiratory monitoring, we introduce the \MC to reconstruct the spectral structure and model temporal coherence to better restore physiological consistency and extract stable respiratory features from encrypted radar signals. To validate its effectiveness, in this section, we compare the respiratory monitoring performance of \MC with several representative feature extraction modules.

Fig.~\ref{ablation_PTN} and Tab.~\ref{tab:MC} illustrate the estimation performance of \MC compared with other algorithms. In Fig.~\ref{ablation}, the horizontal axis denotes the different modules, while the vertical axis represents the corresponding MAE/STD values. It is evident from Fig.~\ref{ablation_PTN} that \MC obtains superior performance, as evidenced by the significantly lower MAE and STD values for respiratory monitoring. By contrast, the estimation errors of BiLSTM, CNN and Transformer models are considerably higher than those of our method. This performance advantage stems from the model's inherent adaptability to restore physiological consistency from encrypted signals. Specifically, as shown in Tab.~\ref{tab:MC}, compared to other models, \MC achieves comparable computational performance. Therefore, it is feasible to extract respiratory features accurately and in real time from encrypted signals without performing decryption or reconstructing the original radar signal.

\subsection{Evaluation on Different Experiment Settings}
In this section, we conduct extensive experiments to comprehensively evaluate the performance of \name under different impact factors, \ie, different distances between the target and radar, respiratory patterns, and durations. 


\subsubsection{Impact of Measurement Distance}
To evaluate the performance of \name at different distances between the target and radar, \ie, 0.5 m, 1.0 m, and 1.5 m, we conduct a series of experiments. In each trial, subjects are instructed to sit still at the designated position. During each experiment, participants are asked to maintain a natural respiratory pattern.

Fig.~\ref{distance} and Tab.~\ref{tab:different} illustrate the impact of varying distances between the target and radar on system performance. The results show that as the distance increases, IRAC decreases significantly, while MAE and STD also decrease slightly. This indicates that greater distances lead to gradual attenuation of the radar signal, thereby weakening the feature strength available for identity recognition and respiration monitoring. Nevertheless, despite this signal degradation, \name still demonstrates high robustness in respiratory monitoring, \ie, even at the farthest distances, the MAE remains within $1.5bpm$. These results suggest that the respiratory features extracted by \name are generalizable in complex distance-varying scenarios.

\subsubsection{Impact of Respiratory Patterns} 
We further conduct experiments under different respiratory patterns to evaluate their impact on system performance. In these scenarios, subjects are instructed to follow three respiratory modes: deep respiration, natural respiration, and irregular respiration. These patterns are deliberately selected to represent a broad spectrum of respiratory conditions that may occur in real-world scenarios. Specifically, to eliminate the interference of other factors, the distance between the radar and the target is fixed at $0.5m$ to ensure a high signal-to-noise ratio and stable measurements.

As shown in Fig.~\ref{pattern} and Tab.~\ref{tab:different}, \name demonstrates strong adaptability and robustness, effectively balancing privacy protection and respiratory monitoring across various respiratory patterns. Specifically, the system achieves the best respiratory monitoring performance during deep respiration, resulting in the lowest MAE and STD of $1.08bpm$ and $0.18bpm$, respectively. At the same time, it provides stronger privacy protection, with IRAC is $34.9\%$. This phenomenon can be attributed to the more pronounced chest wall movements during deep respiration, which enhance the strength and features of respiratory signals, thereby reinforcing both accurate respiration estimation and effective suppression of USI. These results are attributed to the targeted encryption of USI and the robustness of spectral structure reconstruction. Specifically, the former \MB ensures that encryption only affects features related to USI regardless of the breathing pattern, while the latter \MC guarantees robust extraction of respiratory features from perturbed signals.

\subsubsection{Impact of duration}
It is also interesting to know how \name performs when the durations are different, which affect the operations of actual deployment. To evaluate the performance, we carry out five experiments with different durations, \ie, 10s, 15s, 20s, 25s and 30s, which ensures that the data contains at least one complete respiratory cycle.

As illustrated in Fig.~\ref{duration} and Tab.~\ref{tab:different}, the duration exerts a significant influence on the overall system performance. As the duration increases, the MAE, STD, and IRAC metrics all exhibit consistent improvement, indicating that the system is able to learn more discriminative features from longer data sequences. However, when the duration exceeds 20 seconds, all performance indicators begin to plateau. Specifically, the MAE and STD stabilize at approximately $1.2bpm$ and $0.3bpm$, respectively, while IRAC stabilizes at around $32.70\%$, and further increasing the duration does not significantly improve performance. It is worth noting that although increasing duration can improve performance, it also introduces higher higher computational and storage overhead, as well as potential user discomfort during prolonged monitoring. Therefore, setting the monitoring duration to around 20 seconds strikes a suitable balance between the monitoring performance, system efficiency, and user experience.

\section{Conclusion}\label{Conclusion}

In this paper, we investigate the feasibility of using mmWave radar for trusted respiratory monitoring, which could facilitate trusted, high-accuracy, and safe health sensing in collaborative health monitoring systems. To this end, we propose and implement \name, the first trusted respiratory monitoring paradigm leveraging mmWave radar. \name decouples USI from the raw radar signals and perturbs them to achieve data anonymization through the \MA and \MB modules. Furthermore, to ensure accurate respiratory monitoring even under significant waveform variations introduced by encryption perturbations, we design the \MC, a transferred generalized domain-independent network. Based on the principle of Spectral Distribution Alignment, \MC explicitly reconstructs the spectral structure and models temporal coherence to better restore physiological consistency and extract stable respiratory features from encrypted radar signals. Extensive experiments demonstrate that \name effectively balances identity privacy and respiratory monitoring performance, achieving strong USI anonymization while maintaining high accuracy in respiratory monitoring from perturbed waveforms. In addition, a comprehensive study of different impacts (\ie, measurement distance, different breathing patterns, and various durations) further confirms the robustness of our system.

\bibliographystyle{IEEEtran}
\bibliography{bib/con.bib}

@ARTICLE{10632107,
  author={Zhang, Lei and Bao, Rong and Jiahao, Chen and Zhu, Yonghong},
  journal={IEEE Transactions on Consumer Electronics}, 
  title={Smart City Healthcare: Non-Contact Human Respiratory Monitoring With WiFi-CSI}, 
  year={2024},
  volume={70},
  number={3},
  pages={5960-5968},
  doi={10.1109/TCE.2024.3441009}}

@ARTICLE{10414999,
  author={Wang, Chaowei and Wang, Ziye and Guan, Weiwei and Wang, Wenjie and Xu, Lexi and Li, Lihua and Huang, Sai and Wang, Weidong},
  journal={IEEE Transactions on Consumer Electronics}, 
  title={Trustworthy Health Monitoring Based on Distributed Wearable Electronics With Edge Intelligence}, 
  year={2024},
  volume={70},
  number={1},
  pages={2333-2341},
  doi={10.1109/TCE.2024.3358803}}

@ARTICLE{11175575,
  author={Gao, Yuan and Zhang, Chenglong and Chen, Ziyang and Zhang, David},
  journal={IEEE Transactions on Consumer Electronics}, 
  title={Smart Breath Analysis for IoMT Consumer Electronics: Feature Engineering for Health Applications}, 
  year={2025},
  volume={},
  number={},
  pages={1-1},
  doi={10.1109/TCE.2025.3613409}}

@article{liu2025multi,
  title={Multi-user Behavioral Privacy Filtering for mmWave Radar Sensing},
  author={Liu, Xiulong and Liu, Hankai and Zhang, Jiaqi and Xie, Xin and Li, Keqiu},
  journal={IEEE Transactions on Mobile Computing},
  year={2025},
  publisher={IEEE}
}

@article{hua2019cosine,
  title={Cosine-transform-based chaotic system for image encryption},
  author={Hua, Zhongyun and Zhou, Yicong and Huang, Hejiao},
  journal={Information Sciences},
  volume={480},
  pages={403--419},
  year={2019},
  publisher={Elsevier}
}

@ARTICLE{10804189,
  author={Wang, Meng and Huang, Jinyang and Zhang, Xiang and Liu, Zhi and Li, Meng and Zhao, Peng and Yan, Huan and Sun, Xiao and Dong, Mianxiong},
  journal={IEEE Network}, 
  title={Target-Oriented WiFi Sensing for Respiratory Healthcare: From Indiscriminate Perception to In-Area Sensing}, 
  year={2025},
  volume={39},
  number={5},
  pages={201-208},
  doi={10.1109/MNET.2024.3518514}}

@article{zhao2024wi,
  title={Wi-pulmo: Commodity wifi can capture your pulmonary function without mouth clinging},
  author={Zhao, Peng and Huang, Jinyang and Zhang, Xiang and Liu, Zhi and Yan, Huan and Wang, Meng and Zhuang, Guohang and Guo, Yutong and Sun, Xiao and Li, Meng},
  journal={IEEE Internet of Things Journal},
  year={2024},
  publisher={IEEE}
}

@article{huang2023phyfinatt,
  title={Phyfinatt: An undetectable attack framework against phy layer fingerprint-based wifi authentication},
  author={Huang, Jinyang and Liu, Bin and Miao, Chenglin and Zhang, Xiang and Liu, Jiancun and Su, Lu and Liu, Zhi and Gu, Yu},
  journal={IEEE Transactions on Mobile Computing},
  year={2023},
  publisher={IEEE}
}

@article{huang2024keystrokesniffer,
  title={Keystrokesniffer: An off-the-shelf smartphone can eavesdrop on your privacy from anywhere},
  author={Huang, Jinyang and Bai, Jia-Xuan and Zhang, Xiang and Liu, Zhi and Feng, Yuanhao and Liu, Jianchun and Sun, Xiao and Dong, Mianxiong and Li, Meng},
  journal={IEEE Transactions on Information Forensics and Security},
  year={2024},
  publisher={IEEE}
}

@inproceedings{li2022spiralspy,
  title={Spiralspy: Exploring a stealthy and practical covert channel to attack air-gapped computing devices via mmwave sensing},
  author={Li, Zhengxiong and Chen, Baicheng and Chen, Xingyu and Li, Huining and Xu, Chenhan and Lin, Feng and Lu, Chris Xiaoxuan and Ren, Kui and Xu, Wenyao},
  booktitle={The 29th Network and Distributed System Security (NDSS) Symposium 2022},
  year={2022},
  organization={The Internet Society}
}

@inproceedings{zhu2022twlbr,
  title={TWLBR: Multi-human through-wall localization and behavior recognition based on MIMO radar},
  author={Zhu, Dongsheng and Wang, Changlong and Han, Chong and Guo, Jian and Sun, Lijuan},
  booktitle={GLOBECOM 2022-2022 IEEE Global Communications Conference},
  pages={3186--3191},
  year={2022},
  organization={IEEE}
}

@article{hao2024mmwave,
  title={Mmwave-RM: A respiration monitoring and pattern classification system based on mmwave radar},
  author={Hao, Zhanjun and Wang, Yue and Li, Fenfang and Ding, Guozhen and Gao, Yifei},
  journal={Sensors (Basel, Switzerland)},
  volume={24},
  number={13},
  pages={4315},
  year={2024}
}

@ARTICLE{10745119,
  author={Liu, Zhaoyu and Xiong, Yuyong and Tian, Wendi and Gou, Yingjie and Peng, Zhike},
  journal={IEEE Transactions on Microwave Theory and Techniques}, 
  title={Large-Scale Body Movement Cancellation in FMCW Radar Respiration Detection}, 
  year={2024},
  volume={},
  number={},
  pages={1-12},
  doi={10.1109/TMTT.2024.3486205}}

@ARTICLE{9864117,
  author={Islam, Shekh M. M. and Borić-Lubecke, Olga and Lubecke, Victor M.},
  journal={IEEE Transactions on Microwave Theory and Techniques}, 
  title={Identity Authentication in Two-Subject Environments Using Microwave Doppler Radar and Machine Learning Classifiers}, 
  year={2022},
  volume={70},
  number={11},
  pages={5063-5076},
  doi={10.1109/TMTT.2022.3197413}}

@ARTICLE{10931026,
  author={Zhao, Haibo and Ma, Yongtao and Kang, Pengfei and Liu, Kaihua},
  journal={IEEE Transactions on Instrumentation and Measurement}, 
  title={One-to-One Matching of Life Activities and Identity Information in Multitarget Scenarios Using Millimeter-Wave Radar and RFID}, 
  year={2025},
  volume={74},
  number={},
  pages={1-11},
  doi={10.1109/TIM.2025.3552378}}

@inproceedings{yang2020mu,
  title={MU-ID: Multi-user identification through gaits using millimeter wave radios},
  author={Yang, Xin and Liu, Jian and Chen, Yingying and Guo, Xiaonan and Xie, Yucheng},
  booktitle={IEEE INFOCOM 2020-IEEE conference on computer communications},
  pages={2589--2598},
  year={2020},
  organization={IEEE}
}

@inproceedings{shenoy2022rf,
  title={Rf-protect: privacy against device-free human tracking},
  author={Shenoy, Jayanth and Liu, Zikun and Tao, Bill and Kabelac, Zachary and Vasisht, Deepak},
  booktitle={Proceedings of the ACM SIGCOMM 2022 Conference},
  pages={588--600},
  year={2022}
}

@article{lu2025explicit,
  title={Explicit estimation of magnitude and phase spectra in parallel for high-quality speech enhancement},
  author={Lu, Ye-Xin and Ai, Yang and Ling, Zhen-Hua},
  journal={Neural Networks},
  pages={107562},
  year={2025},
  publisher={Elsevier}
}

@article{liu2023application,
  title={Application-oriented privacy filter for mmWave radar},
  author={Liu, Hankai and Liu, Xiulong and Xie, Xin and Tong, Xinyu and Shi, Tuo and Li, Keqiu},
  journal={IEEE Communications Magazine},
  volume={61},
  number={12},
  pages={168--174},
  year={2023},
  publisher={IEEE}
}

@inproceedings{xu2024mmear,
  title={mmEar: Push the limit of COTS mmWave eavesdropping on headphones},
  author={Xu, Xiangyu and Chen, Yu and Ling, Zhen and Lu, Li and Luo, Junzhou and Fu, Xinwen},
  booktitle={IEEE INFOCOM 2024-IEEE Conference on Computer Communications},
  pages={351--360},
  year={2024},
  organization={IEEE}
}

@INPROCEEDINGS{10229095,
  author={Feng, Yiwen and Zhang, Kai and Wang, Chuyu and Xie, Lei and Ning, Jingyi and Chen, Shijia},
  booktitle={IEEE INFOCOM 2023 - IEEE Conference on Computer Communications}, 
  title={mmEavesdropper: Signal Augmentation-based Directional Eavesdropping with mmWave Radar}, 
  year={2023},
  volume={},
  number={},
  pages={1-10},
  doi={10.1109/INFOCOM53939.2023.10229095}}

@article{wang2024real,
  title={Real-time through-wall multiperson 3D pose estimation based on MIMO radar},
  author={Wang, Changlong and Zhu, Dongsheng and Sun, Lijuan and Han, Chong and Guo, Jian},
  journal={IEEE Transactions on Instrumentation and Measurement},
  year={2024},
  publisher={IEEE}
}

@article{qiu2023radar,
  title={Radar 2: Passive Spy Radar Detection and Localization Using COTS mmWave Radar},
  author={Qiu, Yanlong and Zhang, Jiaxi and Chen, Yanjiao and Zhang, Jin and Ji, Bo},
  journal={IEEE Transactions on Information Forensics and Security},
  volume={18},
  pages={2810--2825},
  year={2023},
  publisher={IEEE}
}

@article{zhao2024mn,
  title={MN-UIV: Multimodal neural network enabling user identity verification based on millimeter wave radar},
  author={Zhao, Haibo and Ma, Yongtao and Wang, Xiaofeng},
  journal={IEEE Internet of Things Journal},
  year={2024},
  publisher={IEEE}
}

@article{guo2024millimeter,
  title={Millimeter-Wave Radar-Based Identity Recognition Algorithm Built on Multimodal Fusion},
  author={Guo, Jian and Wei, Jingpeng and Xiang, Yashan and Han, Chong},
  journal={Sensors},
  volume={24},
  number={13},
  pages={4051},
  year={2024},
  publisher={MDPI}
}

@article{hao2025detection,
  title={Detection of vital signs based on millimeter wave radar},
  author={Hao, Zhanjun and Wang, Yue and Li, Fenfang and Ding, Guozhen and Fan, Kai and Gao, Yifei},
  journal={Scientific Reports},
  volume={15},
  number={1},
  pages={28112},
  year={2025},
  publisher={Nature Publishing Group UK London}
}

@ARTICLE{10942423,
  author={Zhang, Zirui and Wu, Weiming and Fu, Ziyu and Han, Zhuo and Zhang, Jinyuan and Sun, Chen and Ma, Dedong and Wang, Cong},
  journal={IEEE Transactions on Instrumentation and Measurement}, 
  title={Real-Time Sleep Apnea Detection and Prediction From Single-Channel Airflow}, 
  year={2025},
  volume={74},
  number={},
  pages={1-9},
  doi={10.1109/TIM.2025.3554899}}

@article{sethuraman2021mywear,
  title={MyWear: A novel smart garment for automatic continuous vital monitoring},
  author={Sethuraman, Sibi C and Kompally, Pranav and Mohanty, Saraju P and Choppali, Uma},
  journal={IEEE Transactions on Consumer Electronics},
  volume={67},
  number={3},
  pages={214--222},
  year={2021},
  publisher={IEEE}
}

@inproceedings{chen2022supervised,
  title={Supervised and self-supervised pretraining based COVID-19 detection using acoustic breathing/cough/speech signals},
  author={Chen, Xing-Yu and Zhu, Qiu-Shi and Zhang, Jie and Dai, Li-Rong},
  booktitle={ICASSP 2022-2022 IEEE International Conference on Acoustics, Speech and Signal Processing (ICASSP)},
  pages={561--565},
  year={2022},
  organization={IEEE}
}

@inproceedings{hu2024m,
  title={M 2-Fi: Multi-person respiration monitoring via handheld WiFi devices},
  author={Hu, Jingyang and Jiang, Hongbo and Zheng, Tianyue and Hu, Jingzhi and Wang, Hongbo and Cao, Hangcheng and Chen, Zhe and Luo, Jun},
  booktitle={IEEE INFOCOM 2024-IEEE Conference on Computer Communications},
  pages={1221--1230},
  year={2024},
  organization={IEEE}
}

@inproceedings{wang2022your,
  title={Your breath doesn't lie: multi-user authentication by sensing respiration using mmwave radar},
  author={Wang, Yao and Gu, Tao and Luan, Tom H and Yu, Yong},
  booktitle={2022 19th Annual IEEE International Conference on Sensing, Communication, and Networking (SECON)},
  pages={64--72},
  year={2022},
  organization={IEEE}
}

@article{song2024finersense,
  title={FinerSense: A Fine-Grained Respiration Sensing System Based on Precise Separation of Wi-Fi Signals},
  author={Song, Wenchao and Wang, Zhu and Guo, Yifan and Sun, Zhuo and Ren, Zhihui and Chen, Chao and Guo, Bin and Yu, Zhiwen and Zhou, Xingshe and Zhang, Daqing},
  journal={IEEE Transactions on Mobile Computing},
  year={2024},
  publisher={IEEE}
}

@ARTICLE{10160149,
  author={Wang, Yingqi and Wang, Zhongqin and Zhang, Jian Andrew and Zhang, Haimin and Xu, Min},
  journal={IEEE Transactions on Mobile Computing}, 
  title={Vital Sign Monitoring in Dynamic Environment via mmWave Radar and Camera Fusion}, 
  year={2024},
  volume={23},
  number={5},
  pages={4163-4180},
  doi={10.1109/TMC.2023.3288850}}

@ARTICLE{10633871,
  author={Hsieh, Chieh-Hsun and Tseng, Po-Hsuan},
  journal={IEEE Transactions on Microwave Theory and Techniques}, 
  title={Multiperson Localization and Vital Signs Estimation Using mmWave MIMO Radar}, 
  year={2025},
  volume={73},
  number={2},
  pages={1222-1234},
  doi={10.1109/TMTT.2024.3434958}}

@article{romano2021non,
  title={Non-contact respiratory monitoring using an RGB camera for real-world applications},
  author={Romano, Chiara and Schena, Emiliano and Silvestri, Sergio and Massaroni, Carlo},
  journal={Sensors},
  volume={21},
  number={15},
  pages={5126},
  year={2021},
  publisher={MDPI}
}

@article{kang2024respiration,
  title={Respiration monitoring of all occupants in a vehicle using time-division multiplexing FMCW radar based on metasurface technology},
  author={Kang, Wei and Zhou, Chenwei and Wu, Wen},
  journal={IEEE Transactions on Microwave Theory and Techniques},
  volume={72},
  number={8},
  pages={4960--4974},
  year={2024},
  publisher={IEEE}
}

@ARTICLE{10965485,
  author={Chrysos, Grigorios G. and Wu, Yongtao and Pascanu, Razvan and Torr, Philip H.S. and Cevher, Volkan},
  journal={IEEE Transactions on Pattern Analysis and Machine Intelligence}, 
  title={Hadamard Product in Deep Learning: Introduction, Advances and Challenges}, 
  year={2025},
  volume={47},
  number={8},
  pages={6531-6549},
  doi={10.1109/TPAMI.2025.3560423}}

@ARTICLE{10456913,
  author={Cosoli, Gloria and Antognoli, Luca and Panni, Luna and Scalise, Lorenzo},
  journal={IEEE Transactions on Instrumentation and Measurement}, 
  title={Indirect Estimation of Breathing Rate Using Wearable Devices}, 
  year={2024},
  volume={73},
  number={},
  pages={1-8},
  doi={10.1109/TIM.2024.3372222}}

@article{xiang2022high,
  title={High-precision vital signs monitoring method using a FMCW millimeter-wave sensor},
  author={Xiang, Mingxu and Ren, Wu and Li, Weiming and Xue, Zhenghui and Jiang, Xinyue},
  journal={Sensors},
  volume={22},
  number={19},
  pages={7543},
  year={2022},
  publisher={MDPI}
}

@article{liu2025gsfl,
  title={GSFL: A Privacy-preserving Grouping-Split Federated Learning Approach in Resource-constrained Edge Computing Scenarios},
  author={Liu, Qi and Wang, Zhilu and Zhou, Xiaokang and Zhang, Yonghong and Liu, Xiaodong and Lin, Haiyang},
  journal={ACM Transactions on Autonomous and Adaptive Systems},
  year={2025},
  publisher={Association for Computing Machinery (ACM)}
}

@ARTICLE{10856209,
  author={Fei, Yiming and Fang, Hao and Yan, Zheng and Qi, Lianyong and Bilal, Muhammad and Li, Yang and Xu, Xiaolong and Zhou, Xiaokang},
  journal={IEEE Transactions on Consumer Electronics}, 
  title={Privacy-Aware Edge Computation Offloading With Federated Learning in Healthcare Consumer Electronics System}, 
  year={2025},
  volume={71},
  number={2},
  pages={4628-4638},
  doi={10.1109/TCE.2025.3535753}}

@ARTICLE{10158936,
  author={Tan, Xudong and Hu, Menghan and Zhai, Guangtao and Zhu, Yan and Li, Wenfang and Zhang, Xiao-Ping},
  journal={IEEE Transactions on Multimedia}, 
  title={Lightweight Video-Based Respiration Rate Detection Algorithm: An Application Case on Intensive Care}, 
  year={2024},
  volume={26},
  number={},
  pages={1761-1775},
  doi={10.1109/TMM.2023.3286994}}

@article{ahmed2023machine,
  title={Machine learning for healthcare radars: Recent progresses in human vital sign measurement and activity recognition},
  author={Ahmed, Shahzad and Cho, Sung Ho},
  journal={IEEE Communications Surveys \& Tutorials},
  volume={26},
  number={1},
  pages={461--495},
  year={2023},
  publisher={IEEE}
}

@ARTICLE{10511074,
  author={Zeng, Yongshen and Yu, Dongfang and Song, Xiaoyan and Wang, Qiqiong and Pan, Liping and Lu, Hongzhou and Wang, Wenjin},
  journal={IEEE Transactions on Instrumentation and Measurement}, 
  title={Camera-Based Cardiorespiratory Monitoring of Preterm Infants in NICU}, 
  year={2024},
  volume={73},
  number={},
  pages={1-13},
  doi={10.1109/TIM.2024.3395314}}

\end{document}